\documentclass[acmsmall,screen,nonacm]{acmart}

\usepackage{amsmath,amsfonts,amsthm}
\usepackage{algorithm}
\usepackage{algpseudocode}
\usepackage{graphicx}
\usepackage{textcomp}
\usepackage{xcolor}
\usepackage{balance}
\usepackage{xspace}
\usepackage{enumitem}
\usepackage{cleveref}
\usepackage[many]{tcolorbox}
\usepackage{subfigure}
\usepackage{booktabs}
\usepackage{siunitx}
\usepackage{colortbl}
\usepackage{multirow}
\usepackage{array}
\usepackage{threeparttable}
\usepackage{pifont}
\usepackage{makecell} %
\usepackage{tabularx} %
\usepackage{wrapfig}

\crefformat{section}{\S#2#1#3} %
\crefformat{subsection}{\S#2#1#3}
\crefformat{subsubsection}{\S#2#1#3}

\newcommand{\mytodocomment}[2]{}

\newcommand{\shuqing}[1]{\mytodopurple{[Shuqing: #1]}}

\newcommand{\todoaftersub}[1]{}

\newcommand{\mytodopurple}[1]{\mytodocomment{purple}{{\sf}~#1}}

\definecolor{mygray}{gray}{.9}

\newcommand{\tool}{\textsc{Orienter}\xspace}
\newcommand{\ivo}{IGE\xspace} %
\newcommand{\ivos}{IGEs\xspace} %

\newcommand{\totalNumberOfDataset}{1,552\xspace}

\newcommand{\toolFullUnderlined}{c\textbf{\underline{O}}ntext-sensitive inte\textbf{\underline{R}}actable GU\textbf{\underline{I}} \textbf{\underline{E}}leme\textbf{\underline{NT}} detection framework for \textbf{\underline{E}}xtended \textbf{\underline{R}}eality apps\xspace}

\newtcolorbox{promptbox}{%
    left=3pt,
    right=3pt,
    top=0pt,
    bottom=0pt,
    boxrule=0mm,
    colback=black!10!white,
    breakable,
    frame empty
}%

\newtcolorbox{examplebox}{%
    left=3pt,
    right=3pt,
    top=0pt,
    bottom=0pt,
    boxrule=0.3mm,
    colframe=black!75!white,
    colback=white,
    breakable
}%

\theoremstyle{definition}

\def\BibTeX{{\rm B\kern-.05em{\sc i\kern-.025em b}\kern-.08em
    T\kern-.1667em\lower.7ex\hbox{E}\kern-.125emX}}
\begin{document}

\title{Grounded GUI Understanding for Vision-Based Spatial Intelligent Agent: Exemplified by Extended Reality Apps}

\author{Shuqing Li}
\orcid{0000-0001-6323-1402}
\affiliation{%
  \institution{The Chinese University of Hong Kong}
  \city{Hong Kong}
  \country{China}
}
\email{sqli21@cse.cuhk.edu.hk}

\author{Binchang Li}
\orcid{0009-0008-5995-4040}
\affiliation{%
  \institution{Harbin Institute of Technology}
  \city{Shenzhen}
  \country{China}
}
\email{24s151125@stu.hit.edu.cn}

\author{Yepang Liu}
\authornote{Yepang Liu is affiliated with both the Research Institute of Trustworthy Autonomous Systems and the Department of Computer Science and Engineering at Southern University of Science and Technology.}
\orcid{0000-0001-8147-8126}
\affiliation{%
  \institution{Southern University of Science and Technology}
  \city{Shenzhen}
  \country{China}
}
\email{liuyp1@sustech.edu.cn}

\author{Cuiyun Gao}
\orcid{0000-0003-4774-2434}
\affiliation{%
  \institution{Harbin Institute of Technology}
  \city{Shenzhen}
  \country{China}
}
\email{gaocuiyun@hit.edu.cn}

\author{Jianping Zhang}
\orcid{0000-0002-8262-9608}
\affiliation{%
  \institution{The Chinese University of Hong Kong}
  \city{Hong Kong}
  \country{China}
}
\email{jpzhang@cse.cuhk.edu.hk}

\author{Shing-Chi Cheung}
\orcid{0000-0002-3508-7172}
\affiliation{%
  \institution{The Hong Kong University of Science and Technology}
  \city{Hong Kong}
  \country{China}
}
\email{scc@cse.ust.hk}

\author{Michael R. Lyu}
\orcid{0000-0002-3666-5798}
\affiliation{%
  \institution{The Chinese University of Hong Kong}
  \city{Hong Kong}
  \country{China}
}
\email{lyu@cse.cuhk.edu.hk}

\begin{abstract}

In recent years, Extended Reality (XR) has emerged as a transformative technology, offering users immersive and interactive experiences across diversified virtual or virtual-real environments. 
Users can interact with XR applications (apps) through interactable GUI elements (\ivos) on the stereoscopic three-dimensional (3D) graphical user interface (GUI). \ivo constitutes the fundamental element of XR GUI, embodying rich semantic information. 
The accurate recognition and precise understanding of these \ivos is instrumental, serving as the foundation of GUI grounding, which can facilitates downstream tasks, including automated XR testing.
A straightforward XR test generator can interact randomly within the app's 3D environment, making it trapped in uninteractable space and resulting in an ineffective and inefficient testing process. 
In contrast, a more intelligent test generator, informed by the accurate locations and semantics of IGEs, can make wiser decisions on interaction targets and orders, forming test sequences that cover more functionalities faster.
The most recent
\ivo detection approaches in SE are designed for 2D mobile apps and typically train a supervised object detection model based on a large-scale manually-labeled GUI dataset, usually
with a pre-defined set of clickable GUI element categories like buttons and spinners.
Such approaches can hardly be
applied to \ivo detection in XR apps, 
due to a multitude of challenges including complexities posed by open-vocabulary and heterogeneous \ivo categories, intricacies of context-sensitive interactability, and the necessities of precise spatial perception and visual-semantic alignment for accurate IGE detection results.
Thus, it is necessary to embark on the \ivo research tailored to XR apps.
In this paper, we propose the first zero-shot 
\toolFullUnderlined
, named \tool.
By imitating human behaviors, \tool observes and understands the semantic contexts of XR app scenes first, before performing the detection.
The detection process is iterated within a feedback-directed validation and reflection loop.
Specifically, \tool contains three components, including (1) \textit{Semantic context comprehension} for capturing the apps' GUI context, (2) \textit{Reflection-directed IGE candidate detection} for identifying and localizing valid GUI elements based on multi-perspective description guided \ivo detection, as well as feedback-directed reflection, and (3) \textit{Context-sensitive interactability classification} which integrates semantic contexts for interactability prediction.
To evaluate our approach and facilitate follow-up research, we spend more than three months constructing the first benchmark dataset which contains 1,552 images from 100 industrial-setting apps on Steam, with 4,470 interactable annotations across 766 semantics categories. 
Extensive experiments on the dataset demonstrate that \tool is more effective than the state-of-the-art GUI element detection approaches (i.e., GPT-4o, OmniParser, YOLO v8, CenterNet2, Faster
R-CNN, UIED, etc.
), surpassing their F1 Score by up to 3.7$\times$ and 121.4$\times$ (1.4$\times$ and 46.2$\times$ on average) in distinguishing the interactibility and semantics of the \ivos, respectively. 
\tool is beneficial\shuqing{No beneficial.} for boosting the performance of automatic testing by isolating the interactable action space from the whole space, regardless of the testing strategies employed.
Experiments demonstrate that \tool-guided testing covers 103.1\% more \ivos with 125.7\% more effective interactions than testing without action space isolation.

\end{abstract}

\received{20 February 2007}
\received[revised]{12 March 2009}
\received[accepted]{5 June 2009}

\maketitle

\section{Introduction}
\label{sec:introduction}

In recent years, Extended Reality (XR) has emerged as a transformative technology, offering users immersive experiences across various virtual or virtual-real environments. 
This technological advancement has catalyzed the development of a myriad of XR apps
~\cite{paper:vr-software-quality}. 
These apps, which span a diverse array of domains, including skill training~\cite{website:xr-application-VirtualSkill}, entertainment~\cite{website:xr-application-xrgames, website:xr-application-xrfilm}, medical procedures~\cite{paper:vr-ar-in-surgery}, and military training~\cite{paper:vr-military-training}, have attracted over 171 million users~\cite{website:vr-users-number}.
This exponential growth underscores a critical need for robust software processes, including development, testing, and maintenance, particularly in high-reliability contexts such as healthcare and military training.
\todoaftersub{XR number!} 

In XR apps, users experience multimodal perceptions through various devices and interact via body movements and gestures.
Among the multimodal perceptions, visual perception obtained via an app's Graphical User Interface (GUI) offers the wealthiest information. 
GUIs in XR apps are often composed of three-dimensional (3D) GUI elements (e.g., images, text, widgets, etc.) 
or real-life objects and users interact with XR apps through \underline{\textbf{I}}nteractable \underline{\textbf{G}}UI \underline{\textbf{E}}lements (short as \textbf{\ivos}).
\todoaftersub{XR.}
\todoaftersub{Citations.}

As an important part of GUI grounding to facilitate downstream tasks, literature has shown that the accurate recognition of \ivos is a cornerstone for many software engineering tasks including automated testing~\cite{paper:conf/fse/XieFXCC20,paper:conf/fse/YeCX0HCXZ21, paper:conf/fse/ChenXXCXZ020/od} and GUI code generation~\cite{paper:conf/ase/NguyenC15, paper:journals/tse/MoranBCBP20, paper:conf/icse/ChenSMXL18}.
For example, Ye et al.~\cite{paper:conf/fse/YeCX0HCXZ21} report that 77\% software experts believe that precise \ivo detection can boost software testing efficiency by at least 50\%.
White et al.~\cite{paper:conf/issta/WhiteFB19} demonstrate that \ivo detection improves branch coverage by 42.5\% compared to random testing. 

\begin{figure}[t!]
\vspace{-1em}
	\centering  
	\subfigure[
 \ivo categories in 
 lab training app~\cite{website:steam-lab-training-app-intro-example} with diverse visual appearances and interaction mechanisms
 ]{
		\label{fig:diversified-object-raw}
        \includegraphics[width=0.22\linewidth]{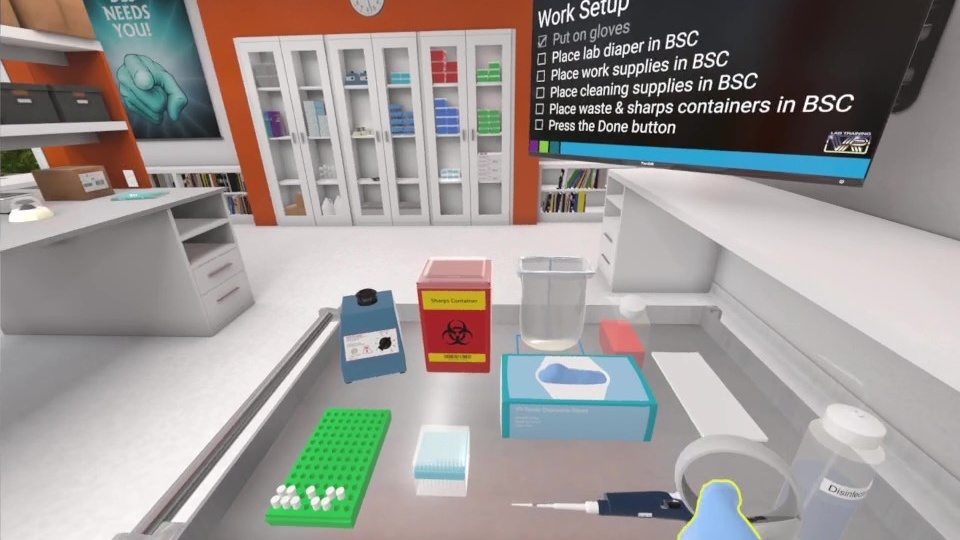}}
	\hspace{1mm}
	\subfigure[\ivo in diversified categories 
 in Figure~\ref{fig:diversified-object-raw} with annotations, dataset annotation requires massive effort]{
		\label{fig:diversified-object-annotated}
        \vspace{-1.8em}
        \includegraphics[width=0.22\linewidth]{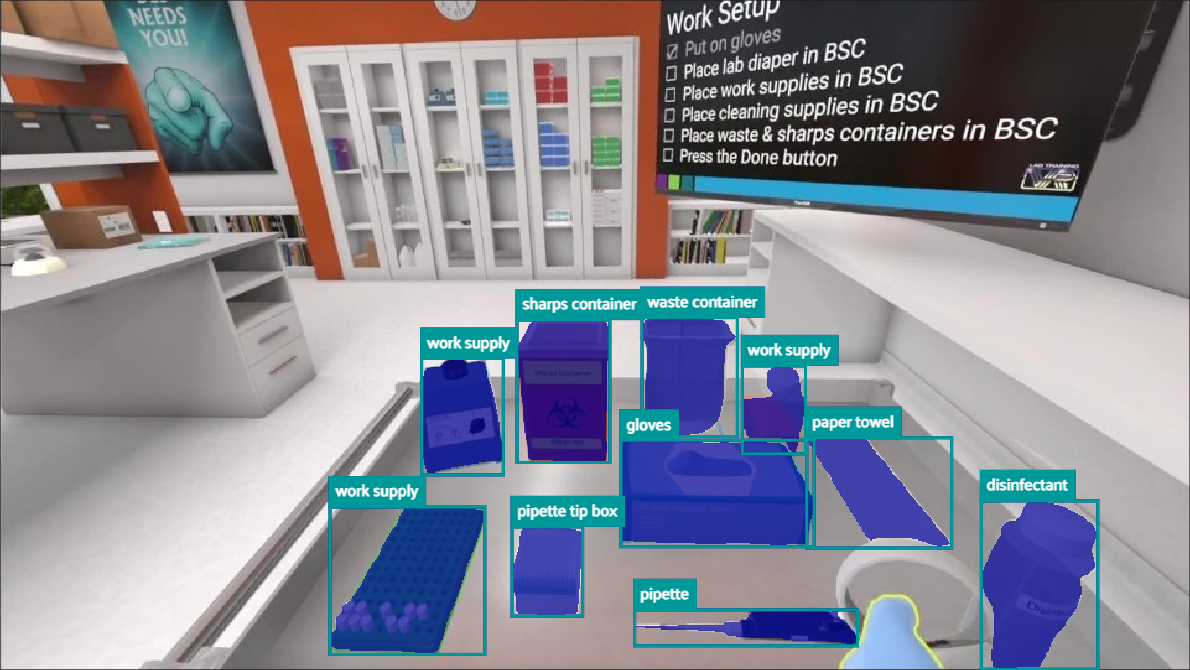}}
	\hspace{1mm}
    \subfigure[An interactable tree in a XR gardening (tree planting) game~\cite{website:steam-planting-app-intro-example}]{
		\label{fig:interactibility-tree-planting-game}
		\includegraphics[width=0.22\linewidth]{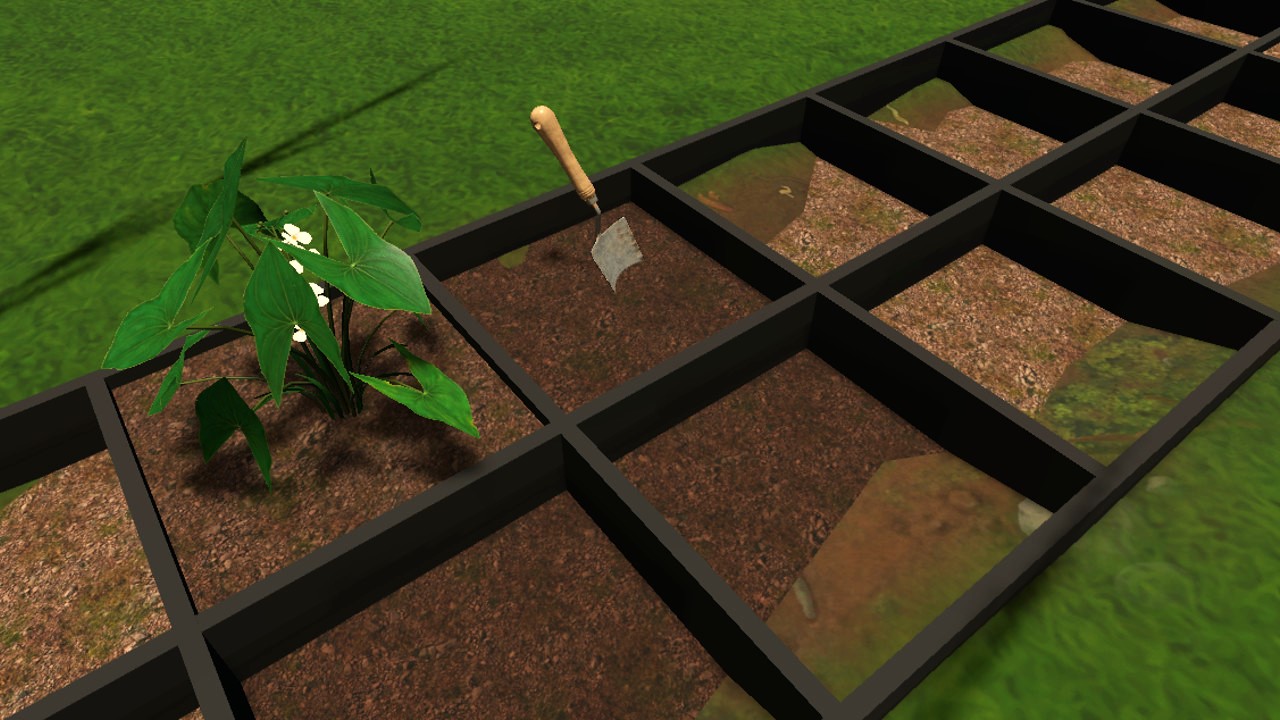}}
	\hspace{1mm}
	\subfigure[Several non-interactable trees in a XR fishing game~\cite{website:steam-lab-fishing-app-intro-example}]{
		\label{fig:interactibility-fishing-game}
		\includegraphics[width=0.22\linewidth]{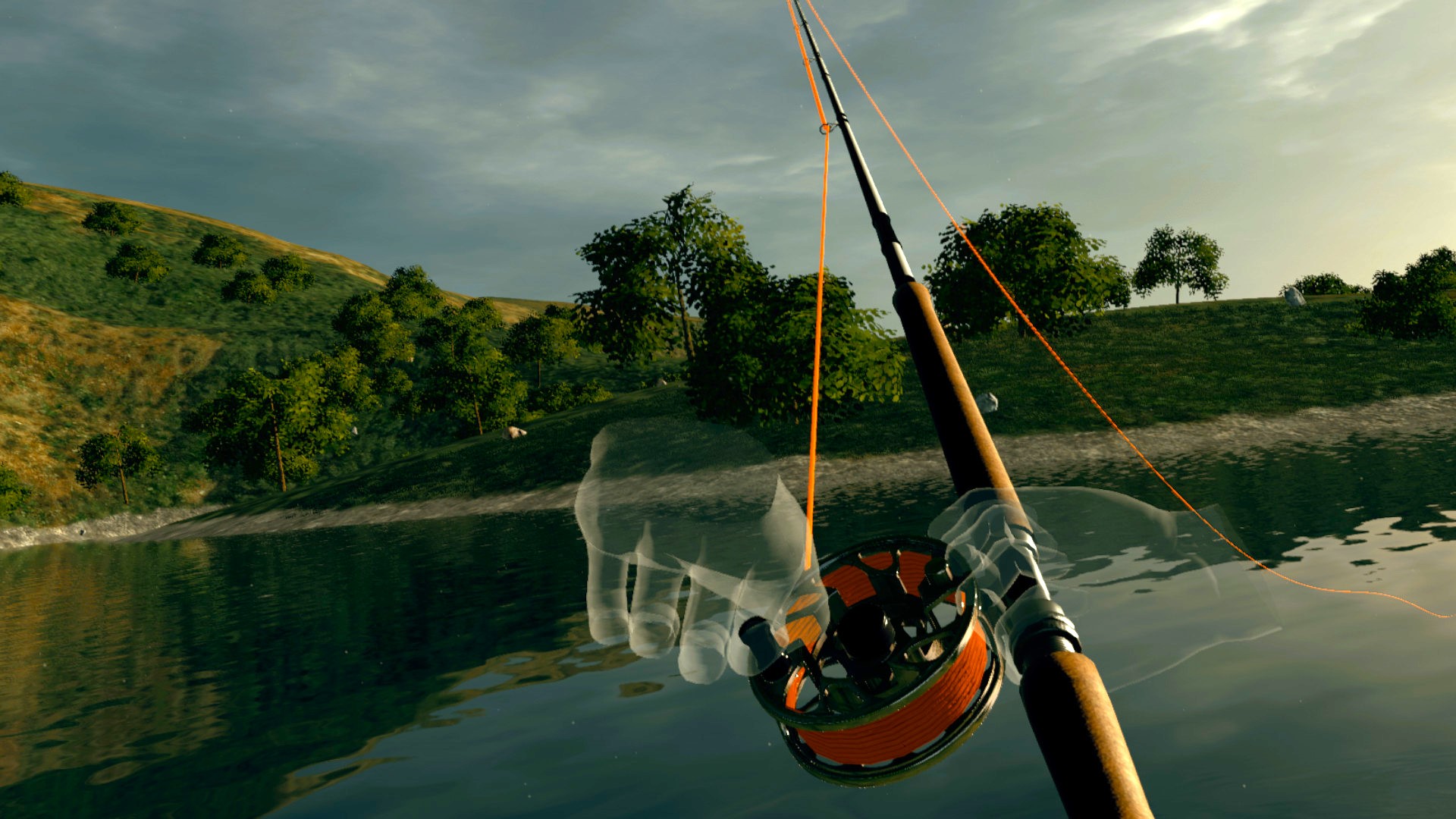}}
    \caption{Examples to demonstrate the challenges of \ivo detection for XR apps~\cite{website:steam-planting-app-intro-example, website:steam-lab-fishing-app-intro-example, website:steam-lab-training-app-intro-example}}
	\label{figs:xr-device}
\end{figure}

Deep learning (DL) based object detection approaches
\cite{paper:journals/pami/fasterrcnn, paper:journals/corr/centernet2, software:yolov8}
have demonstrated promising performance.
With such advancements, recent work has taken steps to explore DL-based \ivo detection approaches in mobile apps and desktop apps~\cite{paper:conf/issta/WhiteFB19, paper:conf/fse/ChenXXCXZ020/od, paper:conf/fse/XieFXCC20, paper:conf/fse/YeCX0HCXZ21}.
For example, Wu et al.~\cite{paper:conf/icse/WuYCXHHMZ23} explore advanced DL-based approaches, like CenterNet2, YOLOv3, and YOLOv5, for IGE detection on mobile apps.
The approaches can be typically divided into three steps: (1) manually summarize a finite set of \ivo categories such as buttons, input boxes, etc., (2) manually label a large dataset, and then (3) train an object detection model on the dataset.
However, such training-based approaches heavily rely on large annotated datasets so that they
can hardly be directly applied to \ivo detection in XR apps.
As shown in Figure~\ref{fig:diversified-object-raw} and Figure~\ref{fig:diversified-object-annotated}, annotating large-scale XR \ivo datasets is challenging due to the unique interaction mechanism of XR apps. 
The data annotation process demands extensive efforts, compounded by XR's complex hardware usage and exhaustive interactions.

\begin{wrapfigure}{r}{0.54\textwidth}
	\centering  
    \vspace{-3em}
	\subfigure[W.r.t heterogeneity]{
		\label{fig:motivating-storm}
		\includegraphics[width=0.37\linewidth]{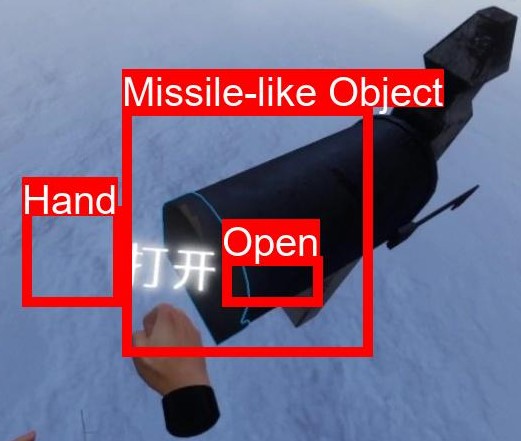}}
	\subfigure[W.r.t. spatial perception]{
		\label{fig:motivating-jobsim}
		\includegraphics[width=0.56\linewidth]{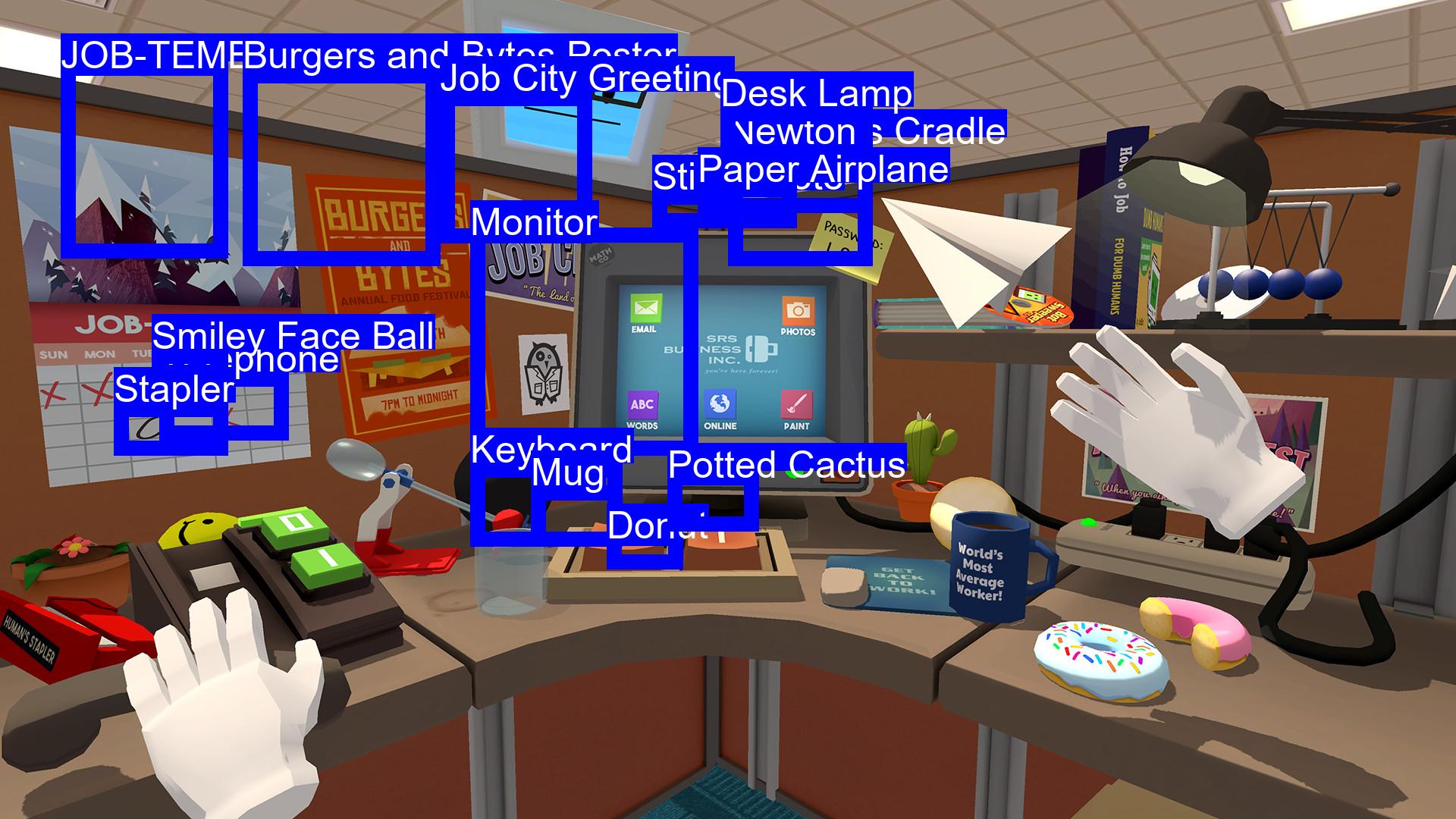}}
    \vspace{-1.3em}
    \caption{LMM gives deviated IGE detection results on VR scenes~\cite{website:steam-game-storm, website:steam-game-jobsim} because of aforementioned challenges}
	\label{figs:motivating-examples}
    \vspace{-1.5em}
\end{wrapfigure}

Recent advancements of pretrained large multimodal models (LMMs)~\cite{paper:journals/corr/gpt4, paper:gemini} 
have shown their remarkable abilities in downstream tasks including image/natural language comprehension, question answering and logical reasoning, without large training sets on specific problems.
Such capabilities provide us with new opportunities to resolve the aforementioned limitation of lacking datasets.
However, our preliminary experiments (detailed in ~\cref{sec:orientor-motivating-example}) reveal that LMMs
suffer from severe spatial semantic hallucinations, tending to generate contextually coherent but factually incorrect or unrealistic \ivos with inaccurate locations and amounts as responses.
These are mainly stemmed from the following challenges of XR \ivo detection problem:

\textbf{\textit{Challenge \#1: Open-vocabulary and heterogeneous GUI element categories}}.
LMM or DL approaches perform well on
detecting
a finite pre-defined set of IGE categories that frequently appear in their training sets and adhere to standard visual appearance patterns (e.g., buttons, sliders, and checkboxes) as well as unified interaction mechanisms (e.g., tapping and long-tapping).
However, XR apps with diverse scenarios contain infinite open-vocabulary categories of IGEs that own different visual appearances and interaction mechanisms (heterogeneity), as shown in Figure~\ref{fig:diversified-object-raw}.
In the stereoscopic 3D XR scene, 
IGEs sometimes occlude each other or are observed from non-front perspectives (e.g., Figure~\ref{fig:motivating-storm}), making the visual appearances of even the same IGE overly diverse and heterogeneous.\shuqing{Example figure.}
Hence, it is hard 
for LMM or DL approaches to identify such open-vocabulary and heterogeneous XR GUI element categories.

\textit{\textbf{Challenge \#2: Context-sensitive interactability.}} 
The interactability of GUI elements in XR apps is highly dependent on the semantic context of the XR app content and scenarios.
As shown in Figure~\ref{fig:interactibility-tree-planting-game} and~\ref{fig:interactibility-fishing-game}, respectively, an object like a tree might be interactable in one app (e.g., a tree-planting game where it needs to be picked up and planted) but non-interactable in another (e.g., as a mere background element in a fishing game).
The IGE
categories that 
LMM or DL approaches perform well when GUI elements with similar appearances behave mostly the same in distinct contexts,
like buttons and checkboxes.
However, the interactability of GUI elements in XR apps is highly dependent on the app-specific and scenario-specific context, which means they cannot be uniformly determined by appearances but require context-sensitive reasoning instead.

\textit{\textbf{Challenge \#3: Accurate IGE detection results require precise spatial perception and visual-semantic alignment.}} 
While capable of processing visual data and generating textual descriptions, LMM struggles with accurate spatial perception. Their ability to outline and label objects
is limited, often resulting in imprecise localization, misaligned bounding boxes, etc. 
This spatial deficiency hinders their applicability in XR environments, where precise IGE localization is necessary.

To tackle the challenges, we propose a \toolFullUnderlined (short as \textbf{\tool}). Instead of directly identifying the IGEs in XR apps, \tool is designed to imitate human behavior
by comprehending the semantic context first before conducting the detection; and the detection process is iterated within a feedback-directed validation and reflection loop (i.e., two levels of ``\textit{looking before leaping}'' in the title). 

Specifically,
\tool mainly contains
three major modules. 
\ding{182} \textit{Semantic context comprehension}: For identifying heterogeneous \ivo categories in an unsupervised and context-sensitive way, we make \tool to understand and combine the global context (app content overview, including genres, storylines, interaction mechanisms, etc.) and local context (the current XR scenario within the user's field of view) of the XR scenes under analysis through in-context reasoning of LMM.
This phase is the foundation of all the subsequent steps.
\ding{183} \textit{Reflection-directed IGE candidate detection}: 
For identifying and localizing valid \ivos, \tool first mines multi-perspective characteristics of GUI elements leveraging LMM, and then uses these characteristic descriptions as guidance to detect corresponding elements. Based on the detection results, \tool further performs feedback-directed reflection to validate and refine \ivo candidates using a feedback loop.
\ding{184} \textit{Context-sensitive interactabilitiy classification}: 
To predict the interactibility of detected IGEs, \tool incorporates the semantic context from the first module and performs chain-of-thought in-context classification.
For facilitating the evaluation, we spend more than three months constructing the first benchmark dataset for \ivo detection in XR apps.
The benchmark dataset consists of 1,552 images from 100 industrial-setting apps on Steam~\cite{website:steam-app-store-vr}, with 4,470 interactable annotations across 766 semantics categories.
We evaluate \tool in terms of predictions on interactability and semantics. \tool outperforms all baselines (GPT-4o~\cite{website:gpt4o}, YOLO v8~\cite{software:yolov8}, CenterNet2~\cite{paper:journals/corr/centernet2}, Faster R-CNN~\cite{paper:journals/pami/fasterrcnn}, UIED~\cite{paper:conf/fse/XieFXCC20}, Xianyu~\cite{paper:conf/fse/ChenXXCXZ020/od}) in almost all metrics, including precision, recall, and F1 score. Specifically, \tool achieves improvements of up to 3.7$\times$ and 121.4$\times$ (1.4$\times$ and 46.2$\times$ on average) on F1 scores in distinguishing the interactability and semantics of \ivos, respectively, demonstrating \tool's strong effectiveness, efficiency, and reliability. %
To illustrate the usefulness of \tool on downstream software engineering (SE) tasks, we further evaluate how effective it boosts the performance of automated testing.
\tool's IGE detection results can be utilized to isolate the interactable action space from the whole input space, regardless of the testing strategies employed.
Experiment results reveal that \tool-guided testing covers 103.1\% more \ivos with 125.7\% more effective interactions than testing without action space isolation.

In summary, we make the following contributions in this paper:
\begin{itemize}[leftmargin=*]
    \item To the best of our knowledge, we are the first to formulate the interactable GUI element (\ivo) detection problem for stereoscopic 3D GUI.
    We leverage the power of large multimodal models in understanding and analyzing GUIs of XR software,
    and propose a zero-shot context-sensitive GUI element detection framework, \tool. \tool novelly undergoes semantic context comprehension, reflection-directed IGE candidate detection, and context-sensitive interactability classification to tackle the challenges of XR IGE detection.
    \item 
    We conduct extensive experiments to verify the effectiveness of the proposed IGE detection framework. The results demonstrate that \tool is more effective than the state-of-the-art GUI element detection approaches.
Experiments show that \tool can facilitate GUI testing.
    \item We construct the first dataset for XR \ivo detection~\cite{website:orientor},  including 1,552 images from 100 industrial-setting apps on Steam, with 4,470 interactable annotations across 766 semantics categories.
\end{itemize}

\section{Background on Extended Reality (XR)}

\textbf{\textit{XR Devices.}}
\label{subsec:xr-devices}
XR technologies offer immersive experiences that seamlessly blend virtual and physical worlds. 
These technologies can be accessed through various device types.
VR typically requires either a PC, a standalone system, or a mobile phone for computational power; 
AR usually overlays digital information onto the physical world, often through smartphones or specialized AR glasses; 
and MR integrates the capabilities of both VR and AR, offering more complex and interactive experiences that merge real and virtual environments.

Typical XR devices include:
(1) \textit{Head-mounted displays} (HMDs), 
which encompass screens that display digital environments and often with audio capabilities.
(2) \textit{Handheld controllers} and \textit{gesture recognition systems}, 
which allows users to navigate and manipulate the virtual environment through physical movements.
(3) \textit{Tracking systems}, 
which includes sensors and cameras, tracks user movements and spatial orientation, enabling a responsive and immersive experience in 3D space.

\begin{wrapfigure}{r}{0.25\textwidth} 
 	\centering 
        \vspace{-1em}
 	\includegraphics[width=\linewidth]{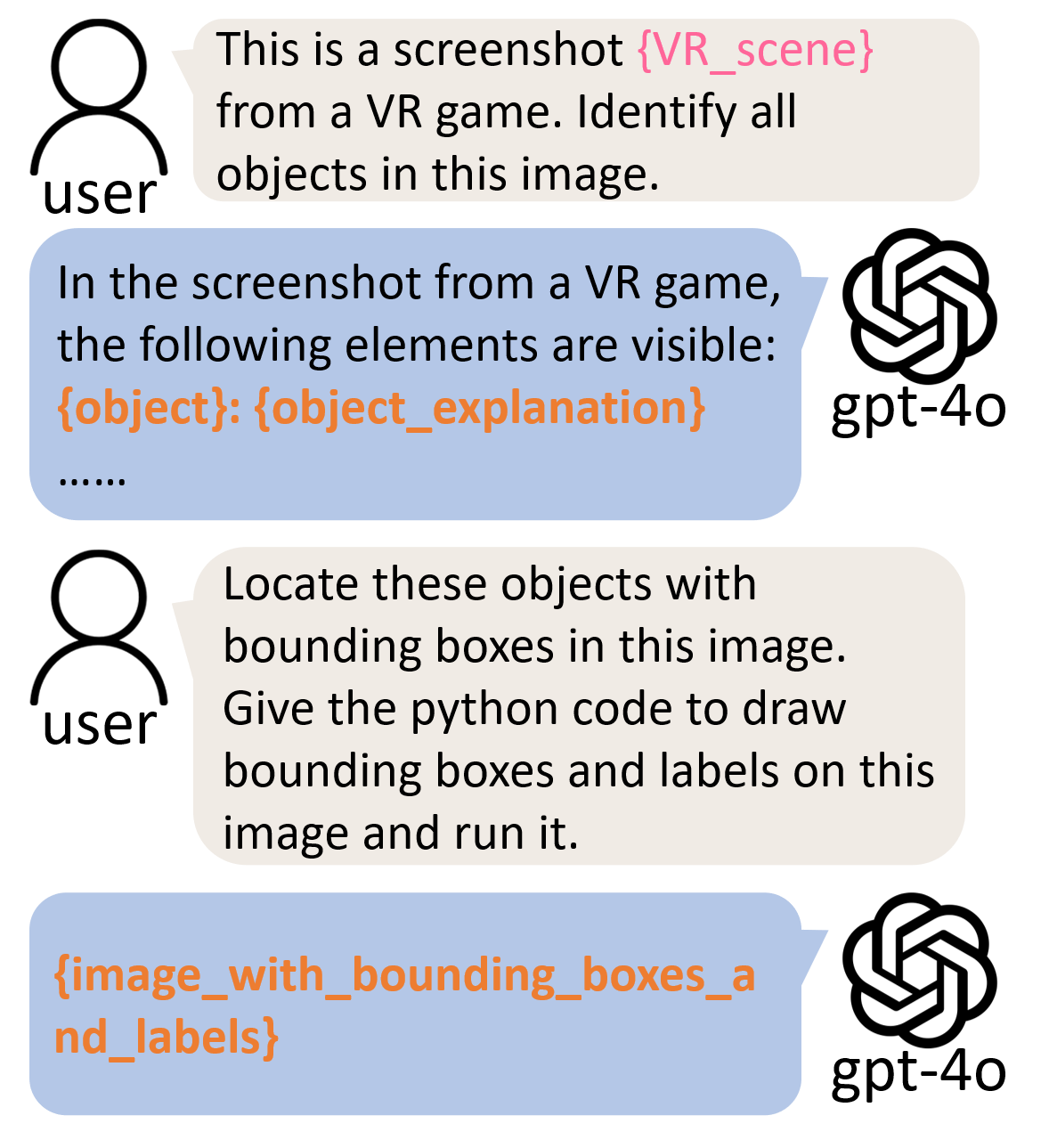} 
        \vspace{-2em}
 	\caption{Prompt template for motivating examples.}
 	\label{fig:motivating-prompt-template}
       \vspace{-1em}
\end{wrapfigure}
\textbf{\textit{XR Interaction.}}
\label{subsec:xr-interaction}
There are various ways in which users can interact with GUI elements. Some examples include:
    (1) \textit{Pressing buttons}. Users can interact with GUI elements by pressing buttons on their devices, such as trigger buttons or joysticks on controllers. 
    (2) \textit{Touching}. Users can touch GUI elements by moving their devices to the location of elements in the virtual world. Some apps extend users' hands in the virtual world with a special object to allow users to interact with elements remotely by making the special object contact with others.
    (3) \textit{Gazing}. Users can interact with GUI elements by directing their gaze to them. 
    Some headsets support eye tracking that enables interaction using only eye movements.
    (4) \textit{Gestures and Movements}. Users can interact with GUI elements through gestures or movements, such as raising or waving hands.
As more methods for human-computer interaction are proposed, the interaction methods are becoming more diverse.

\subsection{Motivating Examples}
\label{sec:orientor-motivating-example}

In this section, we present our motivational study to further demonstrate the technical challenges listed in~\cref{sec:introduction}.
Two authors randomly sample 30 VR apps from the popular Steam VR app store~\cite{website:steam-app-store-vr}, select screenshots from them, and use the prompt in Figure~\ref{fig:motivating-prompt-template} to instruct
LMM GPT-4o~\cite{website:gpt4o} to identify and locate IGEs in the images. %

As shown in figure \ref{fig:motivating-storm}, the mailbox is wrongly recognized as ``Missile-like Object''. It is because the mailbox is viewed from an unusual angle, resulting in a special shape in the view and making it harder for the LMMs to identify semantics. This misunderstanding of GUI elements reveals the LMM's inability to face the heterogeneous GUI elements in VR apps.
For another example, figure \ref{fig:motivating-jobsim}, the bounding boxes in the image either cover only part of the object, locate objects mismatched with the label, or even mark the background. Although the LMM identifies most objects in the image, it fails to locate them well, revealing its insufficient spatial perception. 
To summarize, the LMM alone cannot effectively detect \ivos in highly complex and varied XR apps. To boost the LMM on \ivo detection in XR apps, it is necessary to provide the LMM with more context information and integrate the model with different techniques. Following this idea, we propose \tool to imitate human behavior to enhance the LMM on detecting \ivos in XR apps.

\section{The \tool Approach}

\subsection{Problem Formulation of \ivo Detection}
\label{subsec:problem-formulation}

Before jumping into the details of our approach, we first introduce how we formulate the \ivo detection problem.
The basic IGE detection unit is an individual XR app scenario, 
which lies in the user's field of view, at a specific time.
Unlike mobile application GUIs or web app GUIs, which are entirely 2D, HMD-based XR apps typically render two 2D GUIs for both eyes of users, creating an illusion of depth and making users feel stereoscopic 3D (S3D) sense~\cite{paper:s3d-1, paper:s3d-2}.
These scenes, presented to each eye, are projections of the 3D virtual world onto two 2D "picture planes" (user's eyes), mathematically represented by the function:
$
P_{3D \rightarrow 2D} : UI_{3D} \rightarrow (UI_{left}, UI_{right}),
$
where $UI_{3D}$ represents the 3D scene in the XR environment, and $UI_{left}$ and $UI_{right}$ represent the corresponding 2D projections rendered for the left and right eyes, respectively.
Each one-eye scene can be recovered using the other eye's scene~\cite{paper:stereoid}.
Therefore, the S3D IGEs can be detected by analyzing the 2D projection from a single eye.
This allows us to simplify the XR IGE detection problem to detecting IGEs in the 2D scene of any individual eye.

To facilitate downstream SE tasks such as boosting automated GUI testing, each \ivo is identified not only by its interactability status but also by its location and associated semantic labels, enabling a more fine-grained understanding of the \ivos. 
Let $GE$ represent the set of GUI elements in the XR scene under analysis, i.e., $UI$.
Given that the 3D scene $UI_{3D}$ is projected into 2D for each eye, the right-eye scene can be denoted as:
$
P_{3D \rightarrow 2D}(UI_{right}) : UI_{3D} \rightarrow UI_{right}.
$
The detection function
$
D : GE_{2D} \rightarrow \{Non-interactable, Interactable\}
$
maps each GUI element to a binary interactability determination. The function $D$ operates on the 2D projections of the 3D GUI elements, denoted as:
\[
D(P_{3D \rightarrow 2D}(UI_{right})) : C \times P_{3D \rightarrow 2D}(GE_{3D}) \rightarrow \{Non-interactable, Interactable\} \times B \times S.
\]
Here, $C$ denotes the set of semantic contexts that facilitate context-sensitive analysis.
We use a bounding box $B$ to specify the spatial location of each \ivo, which is the smallest unrotated rectangle containing that \ivo and can be described by its upper-left corner coordinates, width, and height.
Let $S$ denote the set of semantic labels associated with each GUI element, where the labels and their granularity depend on the specific XR scene. 
As presented above, the problem can be simplified to detection on the 2D scene of any individual eye.
If we use the right-eye scene for demonstration, the \ivo detection problem can be formulated as:
\[
D (P_{3D \rightarrow 2D}(UI_{right})) : C \times GE_{2D} \rightarrow \{Non-interactable, Interactable\} \times B \times S.
\]

 \begin{figure*}[t!] 
 	\centering 
 	\includegraphics[width=\textwidth]{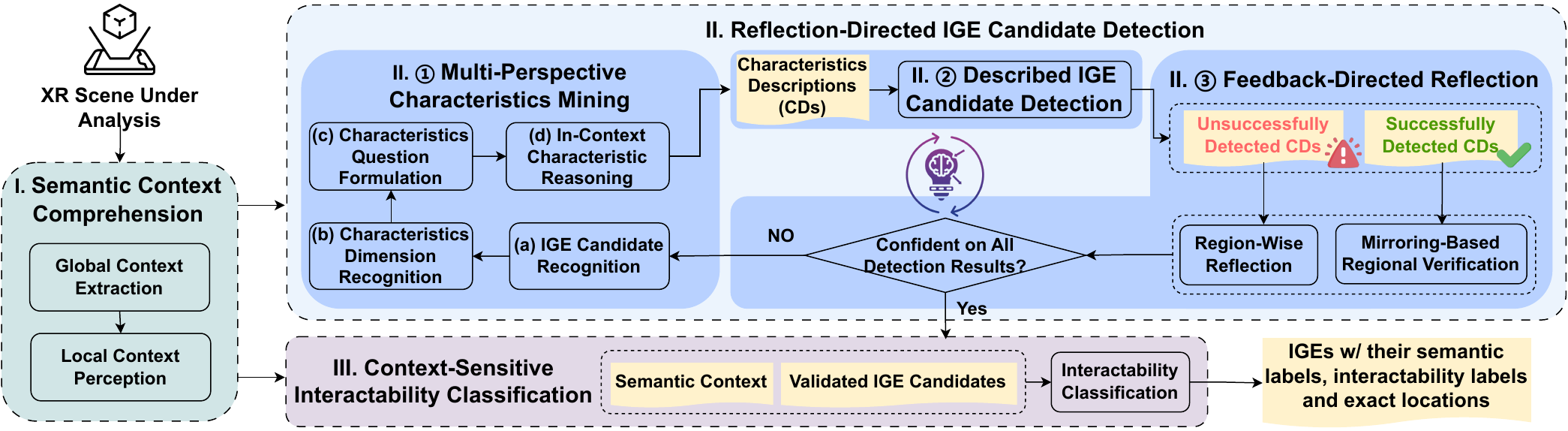} 
        \vspace{-2em}
 	\caption{\shuqing{Testing workflow -- methodology. RQ3 to method. Description} Overview of \tool. The complete prompt templates and example answers can be found on our website. The rounded rectangles in the figure represent modules and submodules. The arrows represent information flow. The light yellow boxes represent diverse forms of data. \todoaftersub{Remove GE. 1.2.3. Context-aware/sensitive. Delete LMM/LLM. Why choose? LMM variants, ablation study, cross-category; data collection. 123, output/input, color different modules}\todoaftersub{Iterations/CLIP,visualize results,}\todoaftersub{Number/Referral}\todoaftersub{Fine tune GDino?}\todoaftersub{MoE?}\todoaftersub{Use traditional CV model for experiments}}
 	\label{fig:overview}
       \vspace{-2em} 
 \end{figure*}

\shuqing{How do we solve the challenges?}
\subsection{Overview of \tool}
\label{subsec:method-overview}

Figure~\ref{fig:overview} illustrates our \toolFullUnderlined, short as \tool.
To address the three challenges (\cref{sec:introduction}), \tool employs a feedback-driven approach that mirrors human behavior, prioritizing semantic comprehension before initiating detection; \tool also enables iterative refinement and validation through a feedback loop. Such designs embody the principle of ``looking before leaping''.
\todoaftersub{from two levels.}

\tool consists of three primary components:
(1) \textit{Semantic Context Comprehension}: To enable unsupervised, context-sensitive IGE detection, \tool first synthesizes both the global context (e.g., app genre, storyline, interaction mechanisms) and the local context (i.e., the user’s current XR field of view). Leveraging the LMM, \tool performs in-context reasoning to establish a comprehensive understanding of the XR scene, forming the basis for subsequent detection.
(2) \textit{Reflection-Guided IGE Candidate Detection}: \tool then mines multi-perspective characteristics of GUI elements, and uses these characteristic descriptions for detecting and locating potential IGEs. 
The framework iteratively refines these candidates through a feedback loop, performing reflection-driven validation to reduce hallucination.
(3) \textit{Context-Sensitive Interactability Classification}: Finally, \tool predicts the interactability of the detected IGE candidates by incorporating the semantic context from the first stage. Using chain-of-thought reasoning within the LMM, \tool classifies interactability without requiring labeled training data, leveraging a knowledge-transfer paradigm from pretrained models in other domains.

\subsection{Module I: Semantic Context Comprehension}
\label{subsec:semantic-context}

The interactability of GUI elements in XR apps is highly dependent on the semantic context of the XR app content and scenarios.
To tackle this challenge, \tool understands and analyzes the target XR app's context beforehand to enable context-sensitive detection.
The semantic context is analyzed comprehensively, from (1) both global (overall app context) and local (current XR scenario) granularities, and (2) both natural language (texts) to vision (images) modalities.
 \begin{wrapfigure}{r}{0.5\textwidth} 
 	\centering 
 	\includegraphics[width=\linewidth]{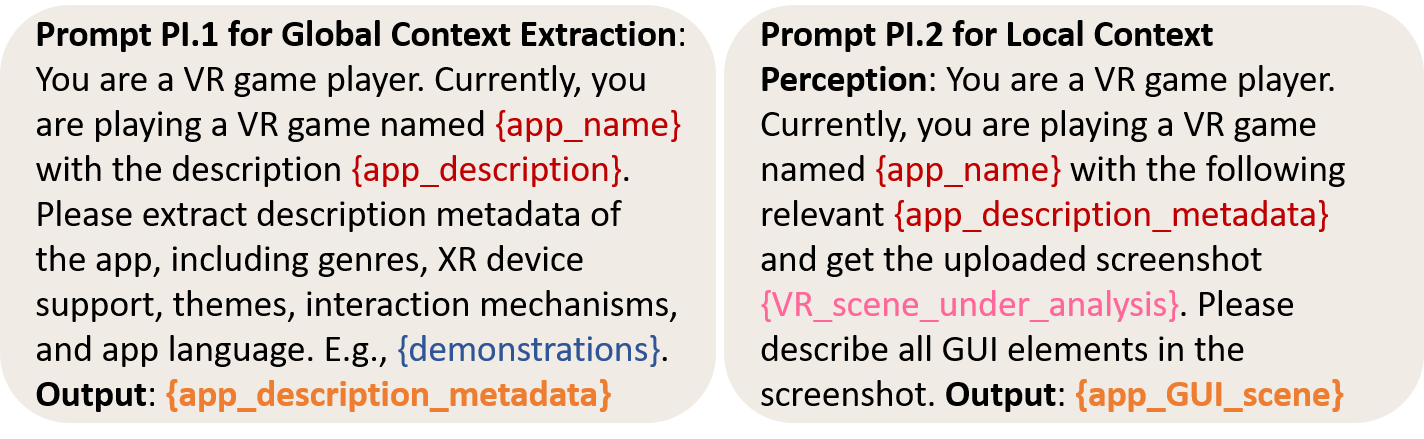} 
        \vspace{-2em}
 	\caption{Prompt templates for Module I\protect\footnotemark}
 	\label{fig:prompts-examples-module-1}
       \vspace{-1em} 
 \end{wrapfigure}

\subsubsection{Global Context Extraction.}
\label{sec:method-global-context}

The global context captures the overall semantic information of the XR app under analysis. 
To find out which global context attributes have influences on IGE interactability, two authors randomly sample 30 real-world VR applications from SteamVR and perform manual inspection following the open coding procedure~\cite{book:open-coding-16}.
At last, we reach a consensus and locate the following key attributes: app names, genres, content themes, VR device supports, ways of gameplay, possible interaction mechanisms, and language information.
Such context can be extracted from the detailed information page on official XR app stores,
but effective extraction can be troublesome due to the free structure natural language format of app details.
Hence, \tool leverages LMM to perform the extraction after the app details are crawled automatically. 
\shuqing{Crawl what content?}
Prompt PI.1 in Figure~\ref{fig:prompts-examples-module-1} shows the prompt template.

\subsubsection{Local Context Perception.}
Local context represents the semantic context within the current XR scenario, i.e., all XR content within the user's current field of view.
Under the extracted global context of app content from~\cref{sec:method-global-context}, \tool further guides LMM to perform in-context visual question answering, to digest and summarize the rendered XR scene as local context.
The guidance of global context ensures that the interpretation of the XR scene is informed by the narrative and purpose of the XR app under analysis.
Specifically, the LMM processes the composite input (screenshot and global context) and outputs a summary of all GUI elements and the background within the scene.
Prompt PI.2 in Figure~\ref{fig:prompts-examples-module-1} shows the prompt template.

\footnotetext{\hypertarget{promptnote} Notes for all prompt templates: Red-colored properties are textual inputs from other modules, pink-colored are image inputs, orange-colored are outputs, and blue-colored are automatically generated information such as CoT demonstrations. We analyze 30 randomly selected SteamVR apps and write CoT reasoning steps and results as demonstrations in the template for LMM. \tool randomly selects three of them as demonstration examples during runtime.}

\subsection{Module II: Reflection-Directed \ivo Candidate Detection}

This module aims at recognizing and localizing IGE candidates within the XR scene, based on the captured semantic contexts from Module I. 
LMM can reveal more semantic hints than traditional methods, making it a better backbone for our approach.
However, LMM tends to generate semantically-incorrect IGEs with inaccurate locations.
This is because XR IGE detection requires precise spatial perception and visual-semantic alignment capabilities, and XR IGE categories are open-vocabulary and heterogeneous (as presented in the challenges and motivational examples).
\tool conquer these challenges through a reflection-directed IGE candidate detection loop:

i). Since LMM has difficulties locating IGE candidates and produces semantic hallucinations, \tool performs IGE candidate detection (\cref{sec:described-ige-candidate-detection}) upon LMM's results, based on a language-model-aligned visual foundation model (VFM).
This VFM detection module (short as VFMD) verifies IGE candidates' existence, reduces hallucinations of incorrect candidates, and locates the exact bounding boxes of existing candidates for further analysis.

ii). Then the VFMD module and LMM work as a chain to conduct IGE detection.
Although VFMD can help reduce hallucinations from LMM, its own analysis process is still error-prone.
To reduce errors in the two components simultaneously, \tool performs iterative validation and refinement alongside the reflection loop (\cref{sec:method-reflection}).
During validation, \tool let another LMM agent, as an advisor (\cref{sec:method-reflection}), compare and rethink all the discrepancies and consistencies between the results from the LMM detector (\cref{sec:method-charac-mining}) and VFMD module.
If any concerns exist for some IGE candidates, the detection chain will run again with the validation comments from the LMM advisor.
The whole process ends when the LMM advisor claims confidence for all detection results.

iii). Inspired by the human cognitive process, when trying to find and locate IGE candidates, using a semantic label only will make the detection process unrobust, e.g., may find the wrong element or find the wrong locations.
Adding more descriptions about the target element (like color, size, shape, relevant locations, etc.) makes the detection more effective and robust.
To boost the effectiveness and robustness of i) and ii), \tool extends a direct LMM detector to a multi-perspective characteristics miner (\cref{sec:method-charac-mining}).

Overall, \tool first mines multi-perspective characteristics of GUI elements leveraging LMM, and then uses these characteristic descriptions as guidance to detect corresponding elements. Based on the detection results, \tool further performs feedback-directed reflection to validate and refine \ivo candidates using a feedback loop.

\subsubsection{\textbf{Multi-Perspective Characteristics Mining}}
\label{sec:method-charac-mining}
\begin{figure}[h!]
 	\centering 
  \vspace{-1em}
 	\includegraphics[width=\columnwidth]{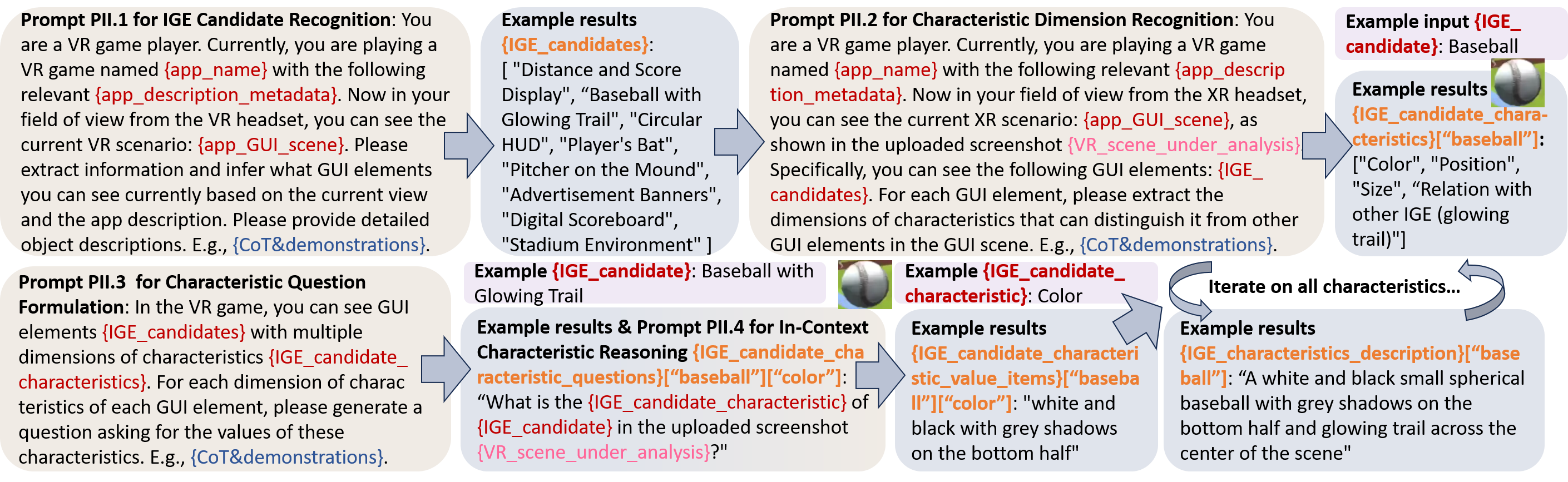} 
        \vspace{-2em}
 	\caption{Prompt templates and examples for Module II.\ding{172}\protect\hyperlink{promptnote}{\textsuperscript{2}}}
 	\label{fig:prompts-examples-module-2.1}
       \vspace{-1em} 
 \end{figure}

In this module, \tool firstly identifies all IGE candidates in the XR scene, and then mines their diverse characteristics from different perspectives (e.g., size, color, shape, relevant location with other objects, etc.).
Then these characteristics will form a characteristics description, which can uniquely identify the corresponding IGE candidate in the present figure in the described detection module.
Figure~\ref{fig:prompts-examples-module-2.1} shows the prompt templates and examples.

  \begin{wrapfigure}{r}{0.35\textwidth} 
 	\centering 
  \vspace{-1em}
 	\includegraphics[width=\linewidth]{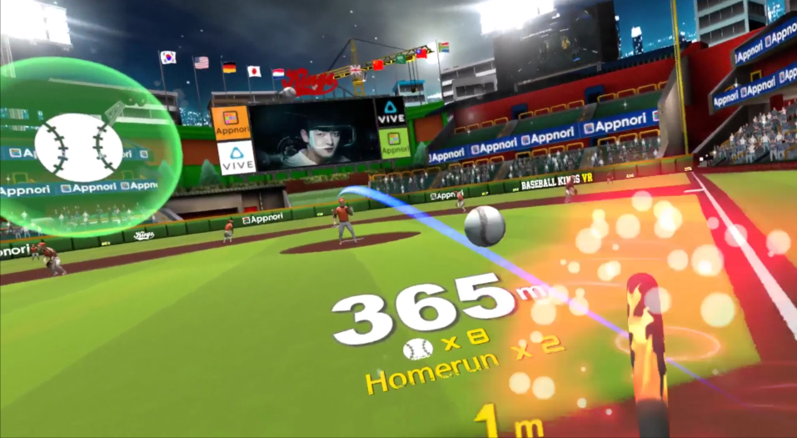} 
        \vspace{-2em}
 	\caption{Example input for Module II.\ding{172}, from the Baseball Kings VR app~\cite{website:steam-baseball-kings}}
 	\label{fig:prompts-example-figure-module-2.1}
       \vspace{-2em} 
 \end{wrapfigure}

\textbf{II.\ding{172}(a): IGE Candidate Recognition} (Prompt PII.1). 
\tool engages in the recognition of IGE candidates within the XR scene. 
With the semantic context from Module I (\cref{subsec:semantic-context}), \tool analyzes the visual semantic cues presented in the environment, finding out all IGE candidates.

\textbf{II.\ding{172}(b): Characteristics Dimension Recognition} (Prompt PII.2). 
Following the identification of IGE candidates, \tool embarks on recognizing the characteristics dimensions of them, which makes them distinguishable from other GUI elements. This involves dissecting the unique characteristics attributes and properties related to the visual appearance,
functionalities, interaction mechanisms, and relative locations of each element. 
\textbf{II.\ding{172}(c) Characteristics Question Formulation} (Prompt PII.3). 
Merging results from the previous two steps, \tool formulates multiple characteristic questions (one question for each dimension of characteristics) for each IGE candidate. 
These questions are tailored to probe the specific appearance, functionalities, relative locations, and interactability dimensions of GUI elements.
\textbf{II.\ding{172}(d) In-Context Characteristics Reasoning} (Prompt PII.4). 
Then \tool conducts characteristics reasoning, querying the LMM to analyze the XR scene and find out the answers for formulated characteristic questions.
This step integrates the local and global contexts previously established, allowing the framework to interpret the responses within the broader narrative and functional scope of the XR application. Through this in-context reasoning, \tool synthesizes a comprehensive understanding of each IGE candidate's characteristics.
\tool then generates \textit{characteristics descriptions} (CDs), which describe the GUI elements based on these characteristics.
Figure~\ref{fig:prompts-examples-module-2.1} shows an example of a characteristics description, ``\textit{A white and black small spherical baseball with grey shadows on the bottom half and glowing trail across the center of the scene}''.

\subsubsection{Described IGE Candidate Detection}
\label{sec:described-ige-candidate-detection}

 \begin{figure}[h!]
 \vspace{-1.5em}
 	\centering 
 	\includegraphics[width=0.7\columnwidth]{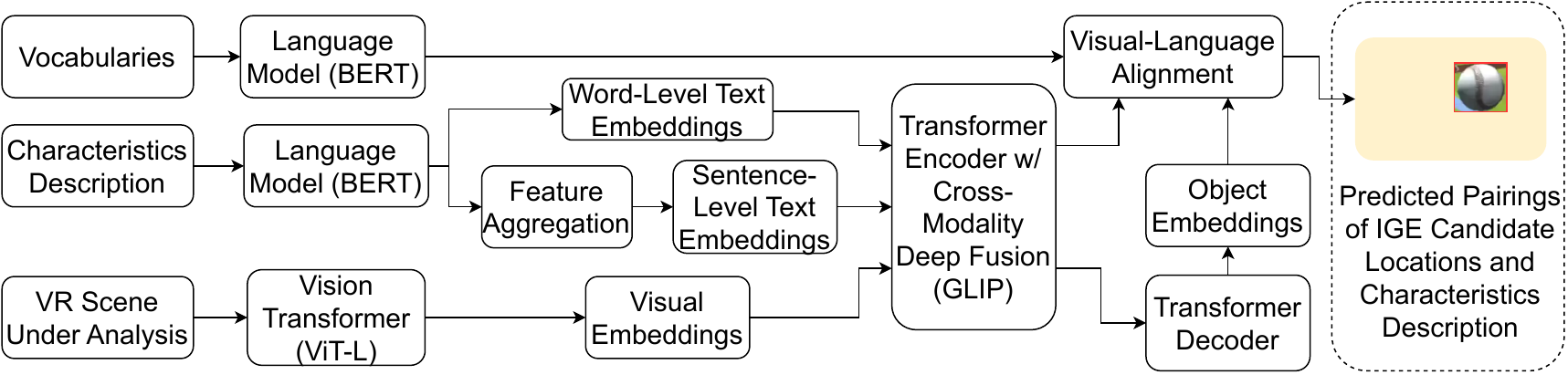} 
        \vspace{-1em}
 	\caption{The framework of the described IGE candidate detection module (VFMD)}
 	\label{fig:ape-overview}
       \vspace{-1.5em} 
 \end{figure}

\tool detects and localizes \ivos using the descriptions of the crafted characteristics generated in the previous step, inspired by Shen's work~\cite{paper:APE}.
\tool employs a language-model-aligned visual foundation model (VFMD).
Figure~\ref{fig:ape-overview} shows the framework structrure.

 \begin{wrapfigure}{r}{0.35\textwidth}
 	\centering 
 	\includegraphics[width=\linewidth]{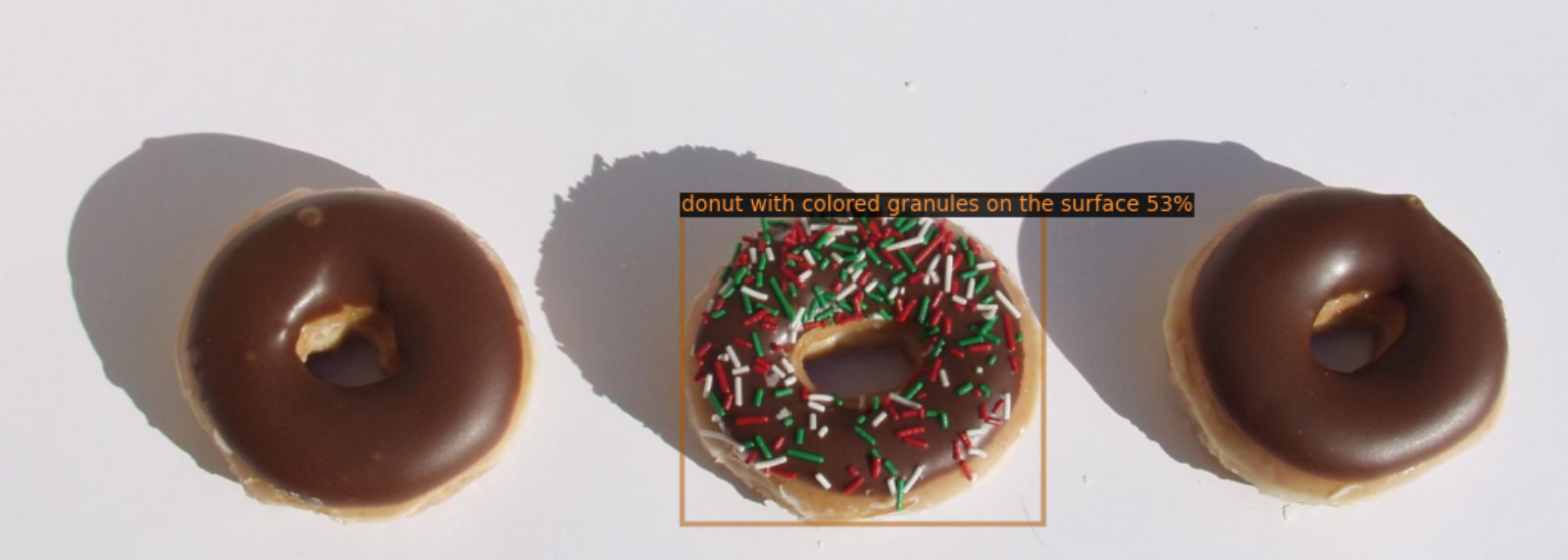} 
        \vspace{-1.3em}
 	\caption{Example: described IGE candidate detection under the CD ``donut with colored granules on the surface'', the donut in the middle is detected.}
 	\label{fig:described-ige-detection-eg-donut}
 \end{wrapfigure}

VFMD consists of a vision backbone for extracting visual features from XR scenes and a language model for generating text embeddings from CDs. Text embedding s are aggregated into sentence-level representations and are fused with visual embeddings using a cross-modality transformer encoder. Object queries conditioned on the text are processed by a transformer decoder, producing object embeddings. A final visual-language alignment module predicts the correct pairings of IGE candidate locations and CDs.
The backbone language model, trained on a vast corpus of data, can generalize to the diverse and unlimited categories of GUI elements present in XR scenarios.
The localization process cross-references (or aligns with) the textual CDs with the visual input from the XR scene. 
By doing, VFMD successfully locates IGE candidates.
Figure~\ref{fig:described-ige-detection-eg-donut} is an example showing the output of this module.

\subsubsection{Feedback-Directed Reflection}
\label{sec:method-reflection}

 \begin{figure}[h!]
 \vspace{-1.5em}
 	\centering 
 	\includegraphics[width=\columnwidth]{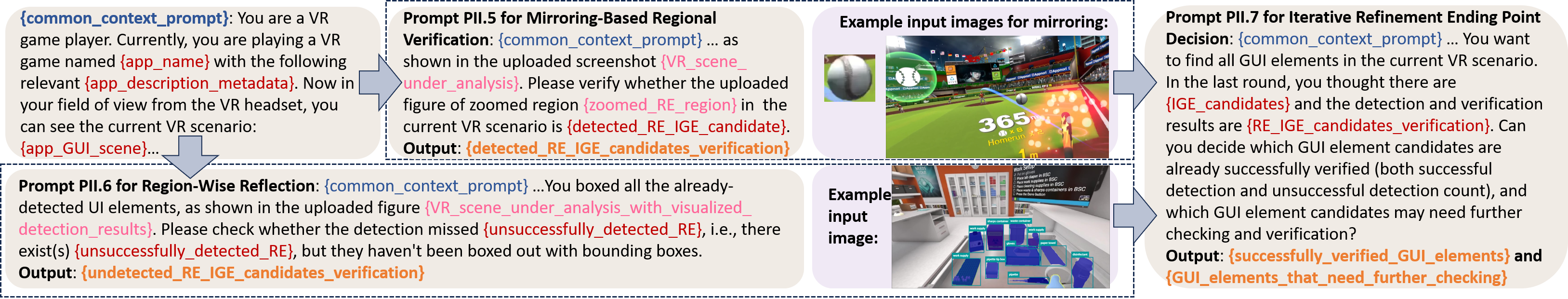} 
        \vspace{-2em}
 	\caption{Prompt templates and examples for Module II.\ding{174}\protect\hyperlink{promptnote}{\textsuperscript{2}}}
 	\label{fig:prompts-examples-module-2.3-reflection}
       \vspace{-1em} 
 \end{figure}

Following previous detection attempts, \tool validates and refines the identified IGE candidates through a feedback-directed reflection loop. 
\tool tries to compare and rethink both the discrepancies and consistencies between the detection results of VFMD module (Module II.\ding{173}) and LMM characteristics miner (Module II.\ding{172}).  
Specifically, \tool distinguishes successfully and unsuccessfully detected CDs, and conducts mirroring-based regional verification and region-wise reflection respectively.
Another LMM agent, as an advisor, analyzes the verification and reflection results and decides whether there exist any concerns for any IGE candidates (Prompt PII.7).
If yes, go back to Multi-Perspective Characteristics Mining module to perform mining again, with the verification results.
\tool iterates through these steps until an LMM advisor decides it reaches sufficient confidence level for all detection results.
Upon reaching a sufficient satisfactory level of confidence in the accuracy of all detected IGE candidates, \tool progresses to the final phase of Context-Sensitive Interactability Classification. 
Figure~\ref{fig:prompts-examples-module-2.3-reflection} shows prompt templates and examples. 

\textbf{Region-Wise Reflection on Unsuccessfully Detected CDs} (Prompt PII.6).
We provide the undetected characteristics descriptions and the visualized \ivo detection results to LMM to validate whether the non-detection is a result of model hallucination or a possible missing detection of the described \ivo candidate detection module.
If LMM decides that it is likely to be a detection miss, \tool will revisit the characteristics mining phase to refine the characteristics descriptions, ensuring a more directed and precise subsequent detection attempt.

\textbf{Mirroring-Based Regional Verification for Successfully Detected CDs} (Prompt PII.5).
Conversely, for successfully detected IGEs, we design a method of regional zooming and mirroring. 
This involves cropping the detected GUI element within its bounding box into a new figure, and \tool lets LMM conduct a comparative analysis against the original XR scene.
Such an approach facilitates a granular verification of the detection's accuracy.

\subsection{Module III: Context-Sensitive Interactability Classification}

 \begin{wrapfigure}{r}{0.38\textwidth} 
 	\centering 
  \vspace{-1.5em}
 	\includegraphics[width=\linewidth]{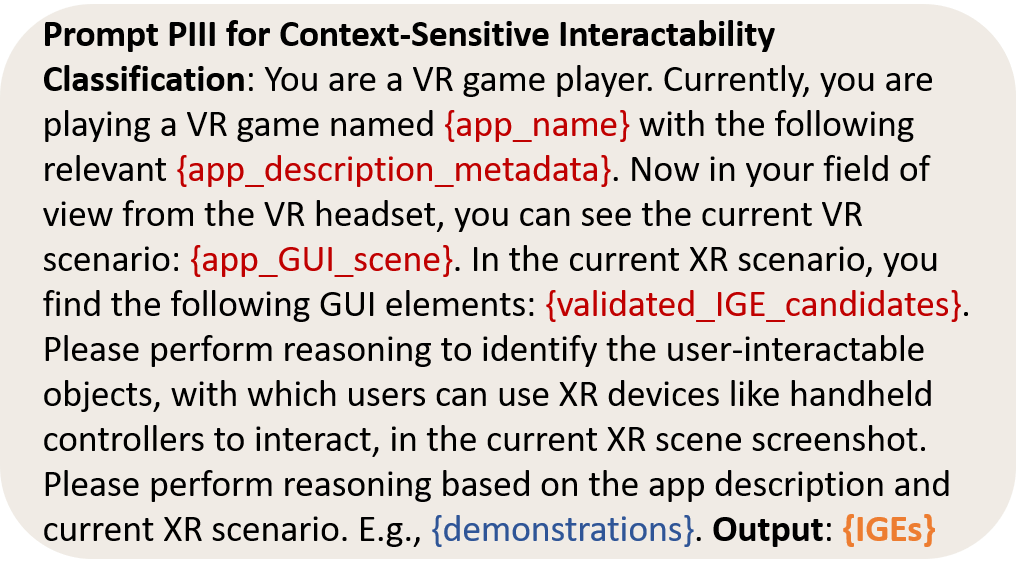} 
        \vspace{-2.5em}
 	\caption{Prompt templates for Module III\protect\hyperlink{promptnote}{\textsuperscript{2}}}
 	\label{fig:prompts-examples-module-3}
 \end{wrapfigure}
After establishing the presence and location of GUI elements in the XR environment through the first two modules, \tool proceeds to analyze their interactivity. 
This module determines which GUI elements in the XR scene can be interacted with.
Such information is useful for many downstream SE tasks, including boosting automated testing performance (detailed in~\cref{sec:rq-usefulness}).

Figure~\ref{fig:prompts-examples-module-3} shows the prompt template PIII.
To classify interactability, \tool guides LMM towards a chain-of-thought in-context reasoning process, within the semantic context captured in Module I.
This process mimics a XR player's thought pattern in distinguishing between GUI elements that are mere visual elements and those that afford interaction. 
For instance, a tree in a tree-planting game (Figure~\ref{fig:interactibility-tree-planting-game}) is interactable, serving as a GUI element to be manipulated by the player, whereas the same tree would be non-interactable background scenery in a fishing game (Figure~\ref{fig:interactibility-fishing-game}).
\tool also provides demonstration examples in the prompt, illustrating the reasoning required for differentiating between interactable and non-interactable GUI elements.
\section{Dataset Construction}\shuqing{Labeling tool. Bounding boxes.}

To the best of our knowledge, there is no existing GUI dataset for \ivo detection in XR apps. Therefore, we build a dataset to verify the usefulness and effectiveness of \tool. 
We recruit a team of 13 annotators, 
spending more than three months in the collection and annotation of GUI images from various XR apps.
These annotators have a minimum of two years of computer science or electrical engineering background, and mostly with experience in video games.
The annotators first interact with GUI elements in various XR apps while their views are recorded. They then select GUI images from these recordings and annotate \ivos. To enhance the quality of the dataset, we train the annotators in the use of XR equipment and the labeling of \ivos, and we ask them to follow several guidelines during data collection and annotation: (1) attempt to identify all \ivos using every possible interaction method in each XR scene, and (2) categorize the semantics of \ivos with appropriate granularity based on their context.

\subsection{Collection of XR Apps}

We first collect XR apps from the Steam app store~\cite{website:steam-app-store-vr}, a comprehensive repository with a wide variety of XR content. 
This collection yielded a total of 4,610 VR-Only apps, which includes 
\textit{software} (official type label on Steam referring to non-games) and \textit{games}.
To obtain a statistically significant sample size with a 95\% confidence level and a 10\% margin of error, 
we randomly sample 102 apps for analysis to cover a broader spectrum of categories, contexts, and interaction paradigms, covering 245 community-generated genres on Steam, demonstrating their diversity and representativeness.

\subsection{Collection of GUI Images from XR Apps}

In this step, the annotators engage with the selected XR apps, interacting with all GUI components within the XR scene in every possible way, as described in~\cref{subsec:xr-interaction}, while recording their stereo view. 
The annotators then select and save GUI images that encompass all XR scenes and GUI elements they explored from the recordings. We crop the right-eye images from the stereo-view videos as only one side of the view is needed, as explained in~\cref{subsec:problem-formulation}. 

\subsection{Annotation of \ivos in GUI Images}

\begin{wrapfigure}{r}{0.45\textwidth}
	\centering  
    \vspace{-5.5em}
	\subfigure[Cooking~\cite{website:steam-game-diner}]{
		\label{fig:fish-in-diner}

        \includegraphics[width=0.4\linewidth]{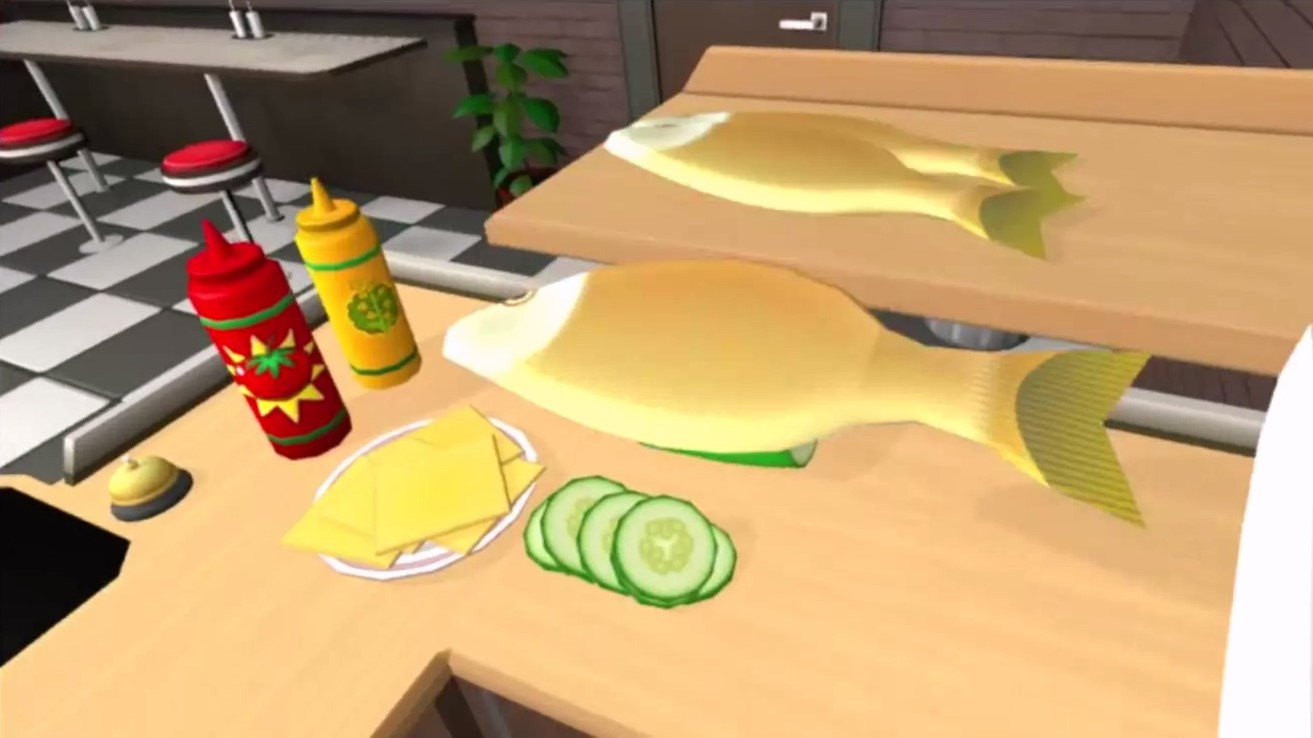}}
	\subfigure[Big-eat-small~\cite{website:steam-game-munch}]{
		\label{fig:fish-in-munch}
		\includegraphics[width=0.4\linewidth]{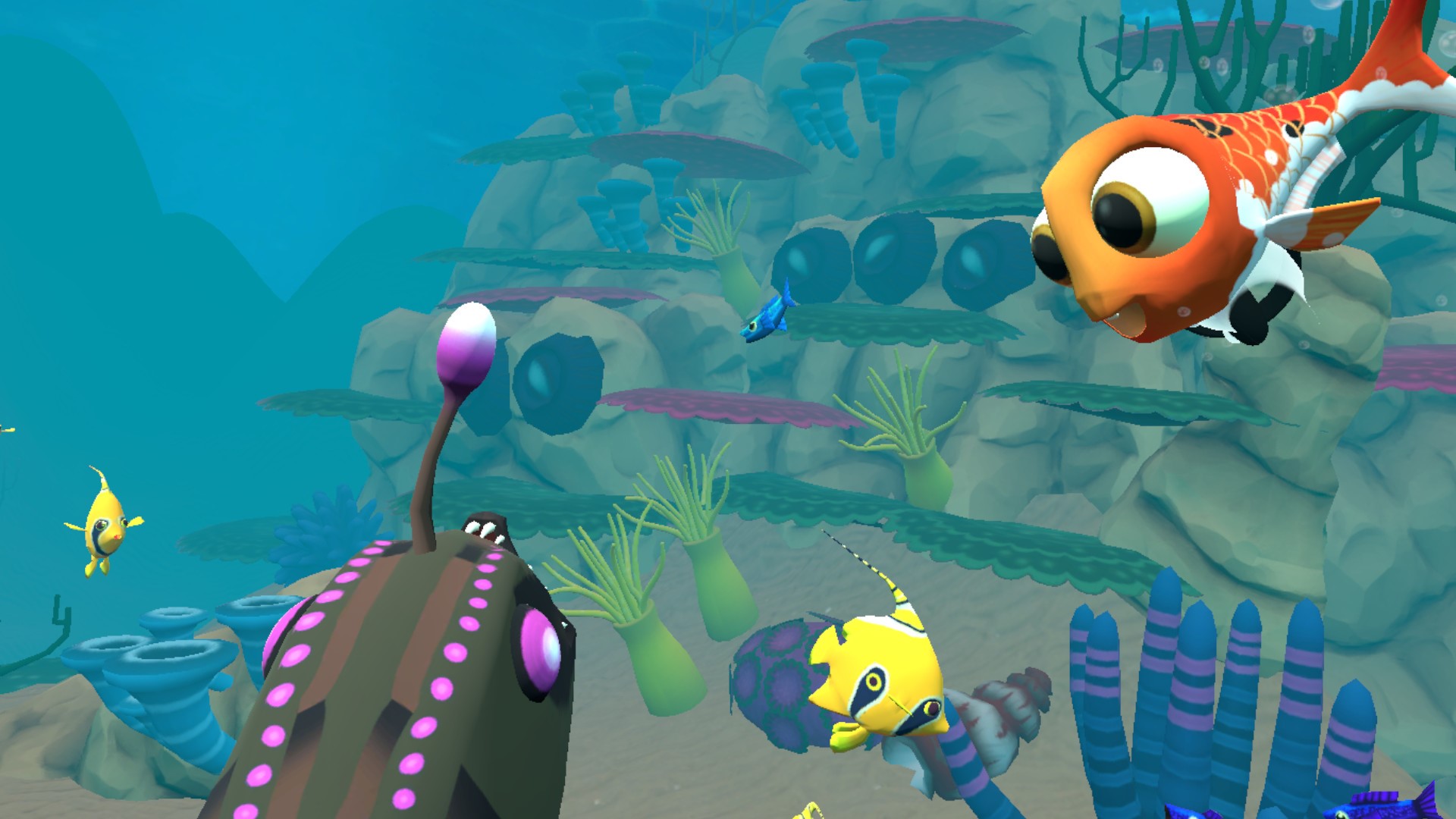}}
    \vspace{-1.5em}
    \caption{Examples to demonstrate varying granularity to best categorize semantics of \ivos.}
	\label{figs:xr-device}
\end{wrapfigure}

The annotators then identify \ivos in GUI images and label their locations with bounding boxes as well as their semantics within the context. As discussed in~\cref{sec:introduction}, the interactability and semantics of GUI elements heavily depend on the diverse contexts of XR apps, making it challenging to categorize all GUI elements with a finite set of predefined categories.

To address this challenge, we ask the annotators to create categories. Specifically, the annotators are instructed to classify \ivos with appropriate granularity based on their experience to better describe the semantics of \ivos in concrete contexts. For example, in the XR game, \textit{VR The Diner Duo}~\cite{website:steam-game-diner}, fish is only one kind of ingredient to make burgers, as shown in \ref{fig:fish-in-diner}. Different individuals of fish are semantically the same in this game. Therefore, a coarse-grained \textit{fish} category is adequate to describe it. However, in the game \textit{Munch VR}~\cite{website:steam-game-munch}, players need to control a fish to eat other smaller fish while avoiding hunted by larger fish, as shown in \ref{fig:fish-in-munch}. In this scenario, fish individuals 
diverge semantically according to size and appearance,
necessitating the introduction of more fine-grained categories such as \textit{large fish} and \textit{small fish} 
to capture their semantic distinctions.

\subsection{Dataset Statistics}

\begin{figure}[h!]
    \vspace{-1.5em}
	\centering  
	\subfigure[\# App in different genres. Genres in different color are used as different set in genre splitting of the dataset introduced in~\cref{subsec:dataset-preparation}]{
		\label{fig:genre-app-count}
		\vspace{-1em}
        \includegraphics[width=0.38\linewidth]{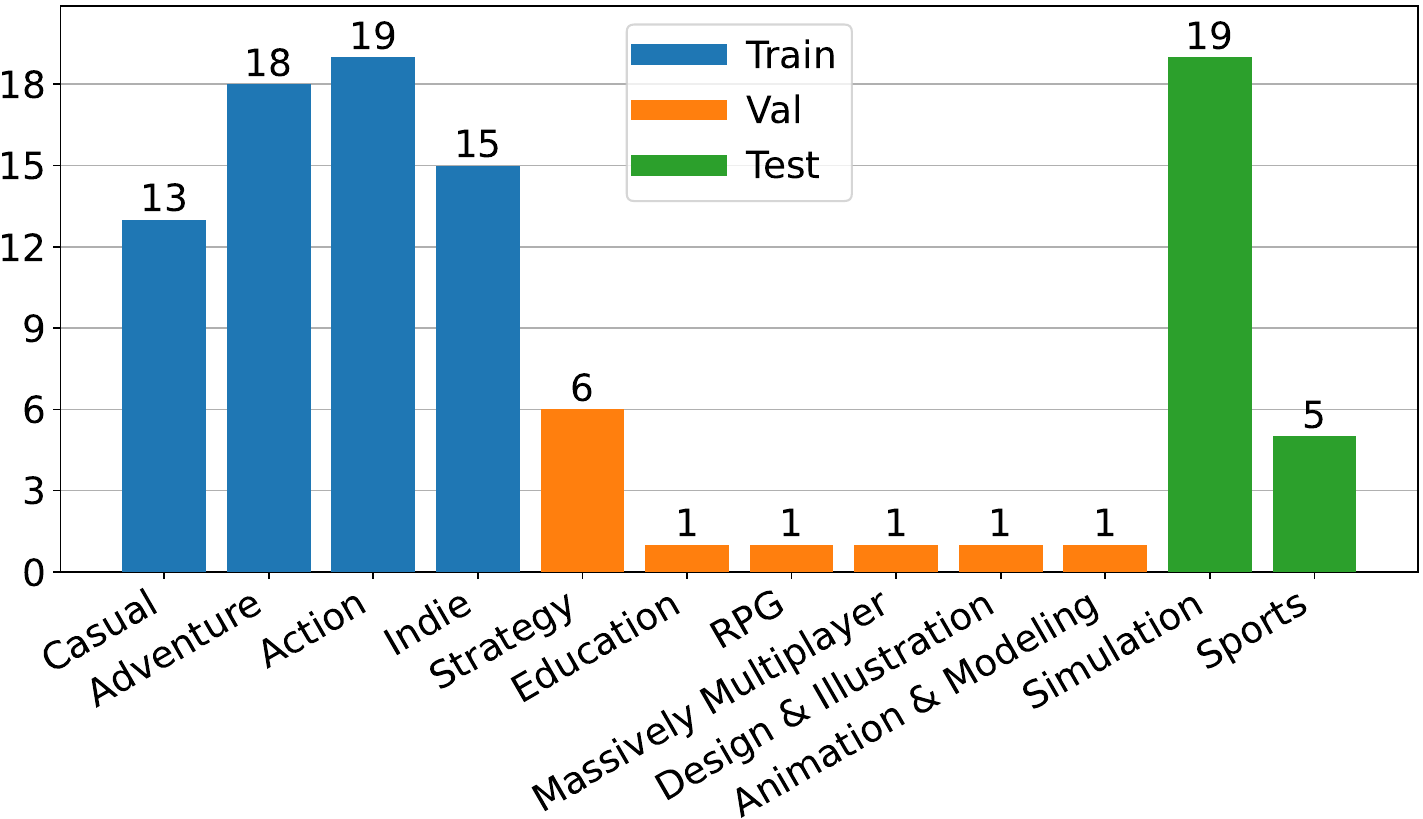}}
	\hspace{1mm}
	\subfigure[\# Annotation in the First 30 \ivo categories associated with the most annotations. The \ivo category "button" which contributes 555 annotations is ommitted]{
		\label{fig:most-common-ige-category}
		\includegraphics[width=0.56\linewidth]{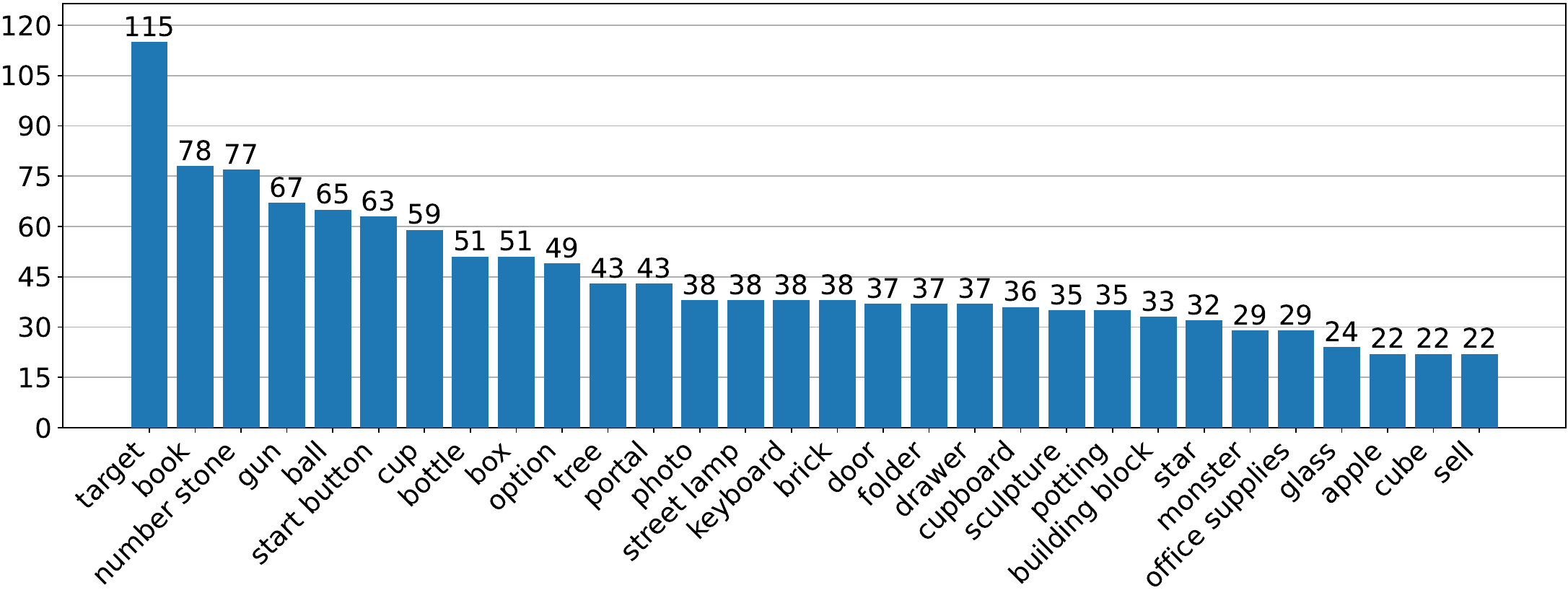}}
    \vspace{-1.5em}
    \caption{Dataset distribution}
	\label{figs:dataset-distribution}
    \vspace{-2em}
\end{figure}

After data cleaning, we construct a dataset consisting of \totalNumberOfDataset images with the size of 960 * 540 from 100 apps, covering 12 official app genres and 245 community-generated app categories on Steam, with 4,470 interactable annotations across 766 semantics categories, highlighting the diversity of our dataset. Annotations are in widely used COCO~\cite{paper:conf/eccv/coco14} format, which uses bounding boxes to mark the locations of \ivos. 
Detailed distribution for apps in app genres and annotations in \ivo's semantics categories are shown in Figure~\ref{fig:genre-app-count} and Figure~\ref{fig:most-common-ige-category}, respectively.

\section{Experiment Design}

\subsection{Research Questions}

In this study, our experiment is designed to answer the following research questions:
\shuqing{Bar charts?}
\begin{itemize}[leftmargin=*, topsep=2pt, itemsep=2pt]
        \item \textbf{RQ1 (IGE detection performance in industrial-setting):} How effective is \tool in \ivo detection on industrial-setting XR apps?
        \begin{itemize}[leftmargin=*, topsep=2pt, itemsep=2pt]
        \item \textbf{RQ1-1 (Interactability)}: How effective is \tool in terms of analyzing interactability? 
        \item \textbf{RQ1-2 (Semantics)}: How effective is \tool in terms of inferring semantics? 
        \item \textbf{RQ1-3 (Context-sensitive interactability)}: How does \tool perform in terms of analyzing context-sensitive interactability, i.e. the same kind of object with different interactability in different contexts?
        \end{itemize}
        \item \textbf{RQ2 (Ablation Study):} How does each component of \tool contributes to its performance? 
        \item \textbf{RQ3 (Effectiveness for improving testing):} How effectively can \tool boost automated testing on XR apps?
        \shuqing{Different directions.}
        \shuqing{Cost, time/money/computational resources}
        
\end{itemize}

\shuqing{LMM/LLM variants}
\shuqing{Cross-category}

\subsection{Baselines}
\label{subsec:baselines}

\begin{wrapfigure}{r}{0.25\textwidth} 
 	\centering 
        \vspace{-2em}
 	\includegraphics[width=\linewidth]{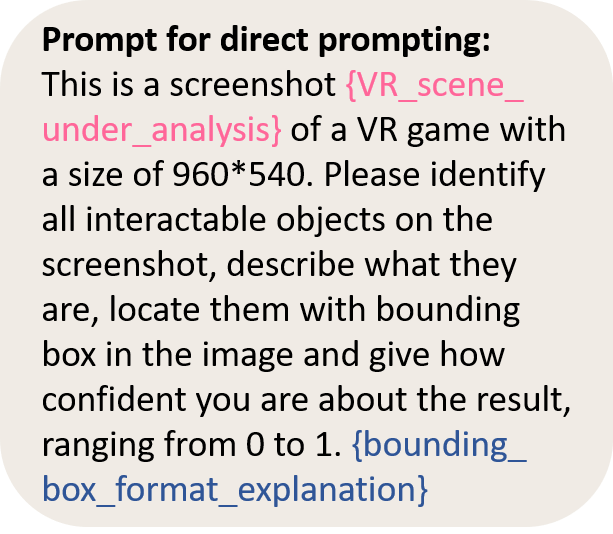} 
        \vspace{-2em}
 	\caption{Direct prompting}
 	\label{fig:direct-prompt}
       \vspace{-2em}
\end{wrapfigure}

\textbf{GPT~\cite{website:gpt4o} \& Gemini~\cite{paper:journals/corr/gemini1.5} \& Claude~\cite{website:claude3.5} } (direct prompting): 
Simply prompt the model with the image and ask it to find the location and semantics of all \ivos.
\textbf{Xianyu}~\cite{paper:conf/fse/ChenXXCXZ020/od} is a UI-to-Code tool leveraging old-fashioned computer vision techniques and OCR. 
\textbf{UIED}~\cite{paper:conf/fse/XieFXCC20} combines old-fashioned methods with deep learning models to detect clickable elements in GUI images.
\textbf{CenterNet2}~\cite{paper:journals/corr/centernet2} is a two-stage object detector that first estimates object probabilities images and then conditionally classify objects.
\textbf{Faster R-CNN}~\cite{paper:journals/pami/fasterrcnn} is a two-stage anchor-based deep learning object detector with a region proposal network and a classification network.
\textbf{YOLO v8}~\cite{software:yolov8} is a one-stage anchor-free detection model that detects and classifies objects simultaneously.
\textbf{OmniParser}~\cite{paper:conf/cvpr/WanSYLC0BY024} combines object detection, OCR, and LMM to parse GUI images into element trees for GUI grounding.

\subsection{Implementation Details and Experimental Setup}

\subsubsection{Implementation of \tool}
For the implementation of \tool, we leverage the most representative and popular LMMs, i.e., gpt-4-vision-preview, gpt-4o-2024-08-06, Claude 3.5 Sonnet, Gemini 1.5 Pro.
For described \ivo detection, we leverage the pretrained model of APE-L\_D~\cite{paper:APE}, which is trained on ten datasets and demonstrates promising results in visual grounding.
We apply post-processing to improve \tool's prediction quality by filtering abnormally large bounding boxes ( $> 90\% $ image size), and applying Non-Maximum Suppression (NMS) which retains only the highest-confidence box among overlapping ones with IoU exceeds 0.7 to reduce duplicates.

\subsubsection{Dataset Preparation}
\label{subsec:dataset-preparation}

To comprehensively answer the RQs, we derive three variations from the original dataset. (1) \textbf{Semantics dataset}: The original dataset testing methods' ability to identify \ivos' semantics, answering RQ1-2. (2) \textbf{Interactability dataset}: All annotations are assigned the category "interactable" for binary classification, testing models' ability to differentiate \ivos, answering RQ1-1. (3) \textbf{Context dataset}: Contains 41 categories randomly sampled from the most common 100 categories, and extra annotations marking their corresponding \textit{non-interactable} objects, testing methods' understanding of context-sensitive \ivos in different contexts, answering RQ1-3.

\begin{wraptable}{r}{0.41\linewidth} 
\centering
\caption{Statistics of different splitting methods. \textit{Ctx.} is short for \textit{Context}.}
\label{tab:split-method-statistics}
\resizebox{\linewidth}{!}{%
\begin{tabular}{@{}ccccc@{}}
\toprule
\textbf{\makecell[c]{Splitting \\ Method}} & \textbf{Set} & \textbf{\makecell[c]{\# Covered \\ App}} & \textbf{\makecell[c]{\# GUI Scene \\ Image}} & \textbf{\# \ivos} \\ \midrule
- & Full & 100 & 1,552 & 4,470 \\ \midrule
\multirow{3}{*}{App} & Train & 60 & 925 & 2,459 \\
 & Valid & 10 & 163 & 574 \\
 & Test & 30 & 464 & 1,437 \\ \midrule
\multirow{3}{*}{Genre} & Train & 65 & 958 & 2,528 \\
 & Valid & 11 & 136 & 610 \\
 & Test & 24 & 458 & 1,332 \\ \midrule
\multirow{5}{*}{Ctx.-sensitive} & Train & 96 & 947 & 2,583 \\
 & Valid & 64 & 158 & 404 \\
 & Test & 53 & 447 & 1,483 \\
 & \makecell[c]{Test \\ (Ctx. dataset)} & 53 & 447 & 1,239 \\ \midrule
$\bigcup$\{App, Genre\} & Test & 46 & 687 & 2,034 \\
\makecell[c]{$\bigcup$\{Ctx.-sensitive, \\App, Genre\}} & Test & 77 & 936 & 2,875 \\ \bottomrule
\end{tabular}
}
\end{wraptable}
We partition the images into training/validation/test sets in a ratio of approximately 6:1:3 using three distinct splitting methods: \textbf{App splitting}, \textbf{Genre splitting}, and \textbf{Context-sensitive splitting}. App splitting randomly allocates images based on their corresponding apps such that images from the same app are placed in the same set. Genre splitting further considers the apps' genres to assess the methods' performance across different app types. Context-sensitive splitting assigns images containing the 41 sampled categories in the context dataset to the test set while randomly distributing the remaining images into the training and validation sets. 

The statistics of each splitting method are illustrated in Table ~\ref{tab:split-method-statistics}. The usage of the datasets and splitting methods will be discussed in~\cref{subsec:experimental-setup} and~\cref{subsec:app-testing-setup}.

\subsubsection{Experimental Setup}
\label{subsec:experimental-setup}

We train baselines on the training set and evaluate all methods including \tool on the test set. For Xianyu~\cite{paper:conf/fse/ChenXXCXZ020/od}, we only use its component detection part to find \ivos.
For UIED~\cite{paper:conf/fse/ChenXXCXZ020/od}, we train the CNN classifier on our dataset to adapt it to our problem. For training on CenterNet2~\cite{paper:journals/corr/centernet2}, Faster R-CNN~\cite{paper:journals/pami/fasterrcnn}, and YOLO v8~\cite{software:yolov8}, we set the batch size to 32 and accordingly adjust the learning rate, apply early stopping with the patience of 20 epoch. Other configurations are the same as their original releases. 
For LMM experiments, we run three times and regard the average results as the final results.

In terms of evaluation, we implement a semantic matching tool to support open vocabulary category matching. We consider two categories semantically match if the cosine similarity of their embedding vectors obtained from the embedding-3 model released by Zhipu AI~\cite{website:zhipuai} exceeds a preset threshold. The first two authors examine the similarity of some typical categories and set the threshold to 0.85. 
We accrodingly modify the official COCO API~\cite{website:coco-api} to support semantic matching.

For RQ1-1, RQ1-2, and RQ2, predictions are evaluated with %
the modified COCO API.
We leverage App splitting and Genre splitting to evaluate each method in RQ1-1 and RQ1-2, then use the union of both splitting's test set to assess the contribution of each components of \tool in RQ2.

Regarding RQ1-3, the baseline methods are trained on the Semantic dataset using Context-sensitive Splitting. Predictions on the test set are then evaluated on the Context dataset with the same splitting. The more \ivos and less non-interactable objects the method detects, the better this method's performance on context-sensitive interactability understanding. 

\subsubsection{XR App Automated Testing Setup}
\label{subsec:app-testing-setup}

To answer RQ3, we simulate a simplified test scenario: given a screenshot of the scene in the XR app under test (AUT), the testing agent attempts to interact with the GUI elements in the scene in a black-box setting. We compare the performance of the testing agent with and without the guidance of \tool using the union of test sets of the 3 splitting as input, regarding the covered 77 apps as AUTs.
The testing process is modeled as incrementally generating points on UI images over time, treating each point as an interaction made by the testing agent with an interval of one minute. We follow the recent works~\cite{paper:conf/issta/PanHWZL20, paper:journals/tosem/GhorbaniJSGM23} to set the testing duration to 60 minutes, which is longer than exploring a XR scene usually requires. We perform 30 testing runs and average the metrics 
for each interaction strategy.

For the non-guided (i.e. random) strategy, the testing agent randomly generates interaction points on the whole image. For the guided strategy, in each attempt, the testing agent generates interaction within the bounding boxes predicted by %
\tool at a probability $p$, or randomly explore the whole UI images at the probability $(1-p)$. The probability $p$ decreases gradually as the test progresses. Suppose the current moment is $t$ minutes and the total duration of the test is $T$ minutes, then $p = 1 - t/T$. This strategy allows the agent to generate effective yet diversified interaction events to cover more \ivos precisely.

\subsection{Evaluation Metrics}
\label{subsec:evaluation-metrics}

For RQ1 and RQ2, we evaluate each method using Precision, Recall, F1-Score, and mAP. The calculation involves computing IoU.

\textbf{IoU}. Intersection over Union (IoU) is calculated as the ratio of two bounding boxes' intersection area to their union area, measuring how well they match. A predefined threshold is usually set, and only pairs of bounding boxes with IoU above this threshold can match. 

\textbf{Precision, Recall, and F1-Score}. 
A predicted bounding box matches the one with the highest IoU among the ground-truth bounding boxes that 
(1) have matching categories, and (2) have IoU exceeding the threshold. 
For RQ1-1, RQ1-2, and RQ2, we consider a predicted box matching a ground-truth box a True Positive (TP); otherwise, a False Positive (FP); ground-truth boxes that fail to match any predicted box are False Negatives (FN). 
For RQ1-2, a predicted box is considered a TP if it matches an interactable ground-truth box, and an FP if it matches a non-interactable one; non-interactable and interactable ground-truth boxes that fail to match any predicted box are considered as TNs and FNs, respectively.
We calculate $Precision = TP/(TP+FP)$, $Recall = TP/(TP+FN)$ and $F1 = 2 \times Precision \times Recall / (Precision + Recall)$. 

\textbf{AP and mAP}. 
We follow COCO API~\cite{website:coco-api} to calculate the Average Precision (AP) by averaging 101 Precision values in increments of 0.01 over a range of Recall values from 0 to 1. The mean AP (mAP) is obtained by averaging AP across categories.

We follow previous works~\cite{paper:conf/icse/Wang22, paper:conf/kbse/WangRM23} to use \ivo Coverage and Effective Interaction Count in RQ3. 

\textbf{\ivo Coverage}.
The \ivo coverage reflects the effectiveness of strategies to find \ivos over time. An \ivo is covered if at least one interaction point falls in the bounding box of that \ivo. \ivo Coverage is calculated as $n_{covered} / n_{all}$, where $n$ represents the number of \ivos.

\textbf{Effective Interaction Count}.
The Effective Interaction Count reflects the efficiency of strategies to make effective interactions by focusing on \ivos only. An interaction point is considered effective if it falls in any of \ivos' bounding boxes.

\section{Results and Analysis}

\subsection{RQ1: IGE detection performance in industrial-setting}

\begin{table*}[t!]
\vspace{-1em}
\centering
\caption{Performance of \tool, DL-based baselines are trained on our dataset\protect\hyperlink{tablenote:explanation}{\textsuperscript{*}}}
\vspace{-1em}
\label{tab:model-performance-is}
\label{tab:model-performance-interactable}
\label{tab:model-performance-semantics}
\resizebox{\columnwidth}{!}{%
\begin{threeparttable}
\begin{tabular}{@{}c|c|ccccc|ccccc|ccccc|ccccc@{}}
\toprule
\multirow{3}{*}{\textbf{Methods}} &  & \multicolumn{10}{c|}{\textbf{Interactability}} & \multicolumn{10}{c}{\textbf{Semantics}} \\
 &  & \multicolumn{5}{c|}{\textbf{App Splitting}} & \multicolumn{5}{c|}{\textbf{Genre Splitting}} & \multicolumn{5}{c|}{\textbf{App Splitting}} & \multicolumn{5}{c}{\textbf{Genre Splitting}} \\
 & \textbf{IoU} & \textbf{0.75} & \textbf{0.80} & \textbf{0.85} & \textbf{0.90} & \textbf{0.95} & \textbf{0.75} & \textbf{0.80} & \textbf{0.85} & \textbf{0.90} & \textbf{0.95} & \textbf{0.75} & \textbf{0.80} & \textbf{0.85} & \textbf{0.90} & \textbf{0.95} & \textbf{0.75} & \textbf{0.80} & \textbf{0.85} & \textbf{0.90} & \textbf{0.95} \\ \midrule
\multirow{4}{*}{\begin{tabular}[c]{@{}c@{}}Faster- \\ RCNN\end{tabular}} & mAP\% & 18.05 & 13.84 & 7.32 & 1.47 & 0.07 & 17.90 & 12.88 & 6.96 & 1.56 & 0.13 & 0.21 & 0.15 & 0.03 & 0.01 & <0.01 & <0.01 & <0.01 & <0.01 & <0.01 & <0.01 \\
& P\% & 33.41 & 32.29 & 23.04 & 11.08 & 2.15 & 33.42 & 29.55 & 20.10 & 8.06 & 1.14 & 0.91 & 0.83 & 0.38 & 0.16 & 0.06 & 0.07 & 0.06 & <0.01 & <0.01 & <0.01 \\
 & R\% & 32.00 & 25.00 & 18.00 & 8.00 & 1.00 & 29.00 & 22.00 & 15.00 & 6.00 & 1.00 & 4.00 & 4.00 & 3.00 & 3.00 & 4.00 & 2.00 & 2.00 & <0.01 & <0.01 & <0.01 \\
 & F1\% & 32.69 & 28.18 & 20.21 & 9.29 & 1.36 & 31.05 & 25.22 & 17.18 & 6.88 & 1.06 & 1.48 & 1.37 & 0.68 & 0.30 & 0.11 & 0.13 & 0.12 & <0.01 & <0.01 & <0.01 \\ \midrule
\multirow{4}{*}{\begin{tabular}[c]{@{}c@{}}Center- \\ Net2\end{tabular}} & mAP\% & 17.79 & 14.98 & 11.73 & 5.91 & 0.93 & 15.63 & 12.53 & 9.07 & 4.23 & 0.45 & 0.20 & 0.18 & 0.15 & 0.05 & <0.01 & 0.22 & 0.19 & 0.05 & 0.02 & <0.01 \\
 & P\% & 32.53 & 28.52 & 26.85 & 18.26 & 4.83 & 28.67 & 25.81 & 27.71 & 16.46 & 4.51 & 0.68 & 0.65 & 0.57 & 0.27 & 0.05 & 0.62 & 0.59 & 0.27 & 0.19 & 0.04 \\
 & R\% & 32.00 & 29.00 & 23.00 & 15.00 & 3.00 & 30.00 & 27.00 & 18.00 & 12.00 & 3.00 & 4.00 & 4.00 & 3.00 & 3.00 & 2.00 & 5.00 & 5.00 & 3.00 & 3.00 & 3.00 \\
 & F1\% & 32.26 & 28.76 & 24.77 & 16.47 & 3.70 & 29.32 & 26.39 & 21.82 & 13.88 & 3.60 & 1.16 & 1.12 & 0.96 & 0.49 & 0.10 & 1.10 & 1.06 & 0.49 & 0.35 & 0.08 \\ \midrule
 \multirow{4}{*}{Yolo v8} & mAP\% & 15.30 & 13.12 & 9.85 & 6.16 & 1.51 & 11.54 & 10.02 & 7.59 & 3.86 & 0.46 & 0.02 & 0.02 & 0.01 & <0.01 & <0.01 & 0.03 & 0.02 & 0.02 & 0.01 & 0.01 \\
 & P\% & 26.58 & 25.38 & 20.63 & 16.65 & 7.31 & 23.92 & 21.38 & 24.22 & 14.55 & 2.68 & 0.21 & 0.20 & 0.19 & 0.05 & 0.01 & 0.29 & 0.28 & 0.27 & 0.25 & 0.21 \\
 & R\% & 26.00 & 23.00 & 20.00 & 14.00 & 3.00 & 23.00 & 22.00 & 15.00 & 11.00 & 3.00 & 4.00 & 4.00 & 4.00 & 1.00 & 3.00 & 3.00 & 3.00 & 3.00 & 3.00 & 3.00 \\
 & F1\% & 26.29 & 24.13 & 20.31 & 15.21 & 4.25 & 23.45 & 21.69 & 18.53 & 12.53 & 2.83 & 0.40 & 0.38 & 0.36 & 0.09 & 0.02 & 0.53 & 0.52 & 0.50 & 0.46 & 0.40 \\ \midrule
 \multirow{4}{*}{OmniParser} & mAP\% & 0.13 & 0.13 & 0.13 & 0.05 & <0.01 & 0.29 & 0.28 & 0.21 & 0.08 & 0.01 & <0.01 & <0.01 & <0.01 & <0.01 & <0.01 & <0.01 & <0.01 & <0.01 & <0.01 & <0.01 \\
 & P\% & 2.48 & 2.44 & 2.22 & 2.37 & 0.29 & 6.19 & 5.90 & 2.93 & 1.76 & 0.96 & <0.01 & <0.01 & <0.01 & <0.01 & <0.01 & <0.01 & <0.01 & <0.01 & <0.01 & <0.01 \\
 & R\% & 2.00 & 2.00 & 2.00 & 1.00 & <0.01 & 3.00 & 3.00 & 3.00 & 2.00 & <0.01 & <0.01 & <0.01 & <0.01 & <0.01 & <0.01 & <0.01 & <0.01 & <0.01 & <0.01 & <0.01 \\
 & F1\% & 2.21 & 2.20 & 2.10 & 1.41 & <0.01 & 4.04 & 3.98 & 2.96 & 1.87 & <0.01 & <0.01 & <0.01 & <0.01 & <0.01 & <0.01 & <0.01 & <0.01 & <0.01 & <0.01 & <0.01 \\ \midrule
 \multirow{4}{*}{\begin{tabular}[c]{@{}c@{}}Gemini 1.5 \\ Pro\end{tabular}} & mAP\% & <0.01 & <0.01 & <0.01 & <0.01 & <0.01 & <0.01 & <0.01 & <0.01 & <0.01 & <0.01 & 0.11 & <0.01 & <0.01 & <0.01 & <0.01 & 0.12 & <0.01 & <0.01 & <0.01 & <0.01 \\
 & P\% & 0.21 & <0.01 & <0.01 & <0.01 & <0.01 & 0.43 & 0.43 & <0.01 & <0.01 & <0.01 & 0.25 & <0.01 & <0.01 & <0.01 & <0.01 & 0.23 & <0.01 & <0.01 & <0.01 & <0.01 \\
 & R\% & <0.01 & <0.01 & <0.01 & <0.01 & <0.01 & <0.01 & <0.01 & <0.01 & <0.01 & <0.01 & 4.00 & <0.01 & <0.01 & <0.01 & <0.01 & \cellcolor{gray!50}50.00 & <0.01 & <0.01 & <0.01 & <0.01 \\
 & F1\% & <0.01 & <0.01 & <0.01 & <0.01 & <0.01 & <0.01 & <0.01 & <0.01 & <0.01 & <0.01 & 0.47 & <0.01 & <0.01 & <0.01 & <0.01 & 0.46 & <0.01 & <0.01 & <0.01 & <0.01 \\ \midrule
\multirow{4}{*}{\begin{tabular}[c]{@{}c@{}}gpt-4o- \\ 2024-08-06\end{tabular}} 
 & mAP\% & 0.01 & <0.01 & <0.01 & <0.01 & <0.01 & 0.01 & <0.01 & <0.01 & <0.01 & <0.01 & <0.01 & <0.01 & <0.01 & <0.01 & <0.01 & <0.01 & <0.01 & <0.01 & <0.01 & <0.01 \\
 & P\% & 0.78 & <0.01 & <0.01 & <0.01 & <0.01 & 0.75 & <0.01 & <0.01 & <0.01 & <0.01 & <0.01 & <0.01 & <0.01 & <0.01 & <0.01 & <0.01 & <0.01 & <0.01 & <0.01 & <0.01 \\
 & R\% & <0.01 & <0.01 & <0.01 & <0.01 & <0.01 & <0.01 & <0.01 & <0.01 & <0.01 & <0.01 & <0.01 & <0.01 & <0.01 & <0.01 & <0.01 & <0.01 & <0.01 & <0.01 & <0.01 & <0.01 \\
 & F1\% & <0.01 & <0.01 & <0.01 & <0.01 & <0.01 & <0.01 & <0.01 & <0.01 & <0.01 & <0.01 & <0.01 & <0.01 & <0.01 & <0.01 & <0.01 & <0.01 & <0.01 & <0.01 & <0.01 & <0.01 \\ \midrule
\multirow{4}{*}{\begin{tabular}[c]{@{}c@{}}\textbf{\tool} \\ \textbf{(gpt-4-vision-} \\ \textbf{preview)}\end{tabular}} & mAP\% & 25.53 & 24.32 & 23.03 & 19.64 & 4.61 & 27.07 & 25.22 & 23.87 & 19.80 & 4.54 & 14.93 & 14.66 & 13.96 & 12.17 & 3.67 & 15.96 & 15.57 & 14.62 & 12.87 & 3.75 \\
 & P\% & 34.79 & 33.52 & 32.32 & 29.31 & 14.92 & 38.42 & 36.70 & 34.97 & 31.40 & 16.79 & 16.06 & 16.05 & 15.17 & 15.27 & 6.53 & 17.11 & 16.85 & 15.78 & 15.93 & 6.14 \\
 & R\% & 40.00 & 39.00 & 38.00 & 34.00 & 12.00 & 40.00 & 38.00 & 37.00 & 32.00 & 11.00 & \cellcolor{gray!50}50.00 & \cellcolor{gray!50}50.00 & \cellcolor{gray!50}50.00 & 33.00 & 20.00 & \cellcolor{gray!50}50.00 & \cellcolor{gray!50}50.00 & \cellcolor{gray!50}50.00 & 33.00 & 20.00 \\
 & F1\% & 37.21 & 36.05 & 34.93 & 31.48 & 13.30 & 39.20 & 37.34 & 35.95 & 31.70 & 13.29 & 24.31 & 24.30 & 23.28 & 20.88 & 9.84 & 25.50 & 25.21 & 23.98 & 21.48 & 9.39 \\ \midrule
\multirow{4}{*}{\begin{tabular}[c]{@{}c@{}}\textbf{\tool} \\ \textbf{(gpt-4o-} \\ \textbf{2024-08-06)}\end{tabular}} & mAP\% & 26.44 & 25.26 & 23.98 & 20.35 & 4.48 & 26.11 & 24.50 & 23.51 & 19.42 & 4.38 & 18.00 & 17.65 & 16.91 & 14.28 & 5.75 & 19.45 & 18.97 & 18.07 & 15.40 & 5.68 \\
 & P\% & 34.63 & 34.89 & 35.43 & 31.64 & 18.57 & 37.91 & 36.68 & 36.71 & 33.24 & 12.91 & 19.32 & 18.97 & 18.28 & 15.37 & 6.87 & 20.35 & 19.81 & 19.01 & 16.19 & 7.36 \\
 & R\% & \cellcolor{gray!50}42.00 & 39.00 & 36.00 & 34.00 & 11.00 & 41.00 & 39.00 & 37.00 & 33.00 & 14.00 & \cellcolor{gray!50}50.00 & \cellcolor{gray!50}50.00 & \cellcolor{gray!50}50.00 & \cellcolor{gray!50}50.00 & \cellcolor{gray!50}42.00 & \cellcolor{gray!50}50.00 & \cellcolor{gray!50}50.00 & \cellcolor{gray!50}50.00 & \cellcolor{gray!50}50.00 & 25.00 \\
 & F1\% & 37.96 & 36.83 & 35.71 & 32.78 & 13.81 & 39.39 & 37.80 & 36.85 & 33.12 & 13.43 & 27.87 & 27.50 & 26.77 & 23.52 & 11.81 & 28.93 & 28.38 & 27.55 & 24.46 & 11.37 \\ \midrule
\multirow{4}{*}{\begin{tabular}[c]{@{}c@{}}\textbf{\tool} \\ \textbf{(Claude 3.5} \\ \textbf{Sonnet)}\end{tabular}} & mAP\% & 23.99 & 22.90 & 21.54 & 18.46 & 3.90 & 25.50 & 23.96 & 22.21 & 18.49 & 3.84 & 16.14 & 15.79 & 15.17 & 12.73 & 5.01 & 19.01 & 18.51 & 17.78 & 15.00 & 6.67 \\
 & P\% & 31.73 & 37.52 & 36.79 & 33.56 & 16.30 & 37.97 & 40.15 & 39.22 & 35.86 & 14.51 & 16.63 & 16.28 & 15.83 & 13.35 & 7.64 & 19.68 & 19.11 & 18.53 & 15.75 & \cellcolor{gray!50}9.87 \\
 & R\% & 40.00 & 32.00 & 31.00 & 29.00 & 11.00 & 37.00 & 33.00 & 32.00 & 29.00 & 13.00 & \cellcolor{gray!50}50.00 & \cellcolor{gray!50}50.00 & \cellcolor{gray!50}50.00 & \cellcolor{gray!50}50.00 & 20.00 & \cellcolor{gray!50}50.00 & \cellcolor{gray!50}50.00 & \cellcolor{gray!50}50.00 & \cellcolor{gray!50}50.00 & 20.00 \\
 & F1\% & 35.39 & 34.54 & 33.65 & 31.11 & 13.14 & 37.48 & 36.23 & 35.24 & 32.07 & 13.71 & 24.95 & 24.56 & 24.04 & 21.07 & 11.06 & 28.24 & 27.66 & 27.04 & 23.96 & 13.22 \\ \midrule
\multirow{4}{*}{\begin{tabular}[c]{@{}c@{}}\textbf{\tool} \\ \textbf{(Gemini 1.5} \\ \textbf{Pro)}\end{tabular}} & mAP\% & \cellcolor{gray!50}32.71 & \cellcolor{gray!50}31.27 & \cellcolor{gray!50}30.51 & \cellcolor{gray!50}26.33 & \cellcolor{gray!50}5.88 & \cellcolor{gray!50}35.03 & \cellcolor{gray!50}32.59 & \cellcolor{gray!50}30.82 & \cellcolor{gray!50}25.08 & \cellcolor{gray!50}5.64 & \cellcolor{gray!50}20.69 & \cellcolor{gray!50}20.35 & \cellcolor{gray!50}20.19 & \cellcolor{gray!50}17.77 & \cellcolor{gray!50}7.16 & \cellcolor{gray!50}22.12 & \cellcolor{gray!50}21.44 & \cellcolor{gray!50}21.17 & \cellcolor{gray!50}18.45 & \cellcolor{gray!50}7.41 \\
 & P\% & \cellcolor{gray!50}46.42 & \cellcolor{gray!50}45.30 & \cellcolor{gray!50}44.94 & \cellcolor{gray!50}41.64 & \cellcolor{gray!50}21.46 & \cellcolor{gray!50}49.45 & \cellcolor{gray!50}47.52 & \cellcolor{gray!50}46.42 & \cellcolor{gray!50}41.94 & \cellcolor{gray!50}19.02 & \cellcolor{gray!50}22.64 & \cellcolor{gray!50}22.29 & \cellcolor{gray!50}22.16 & \cellcolor{gray!50}19.39 & \cellcolor{gray!50}9.66 & \cellcolor{gray!50}23.72 & \cellcolor{gray!50}22.86 & \cellcolor{gray!50}22.64 & \cellcolor{gray!50}19.58 & 8.35 \\
 & R\% & \cellcolor{gray!50}42.00 & \cellcolor{gray!50}41.00 & \cellcolor{gray!50}40.00 & \cellcolor{gray!50}37.00 & \cellcolor{gray!50}14.00 & \cellcolor{gray!50}43.00 & \cellcolor{gray!50}42.00 & \cellcolor{gray!50}41.00 & \cellcolor{gray!50}37.00 & \cellcolor{gray!50}15.00 & \cellcolor{gray!50}50.00 & \cellcolor{gray!50}50.00 & \cellcolor{gray!50}50.00 & \cellcolor{gray!50}50.00 & 25.00 & \cellcolor{gray!50}50.00 & \cellcolor{gray!50}50.00 & \cellcolor{gray!50}50.00 & \cellcolor{gray!50}50.00 & \cellcolor{gray!50}50.00 \\
 & F1\% & \cellcolor{gray!50}44.10 & \cellcolor{gray!50}43.04 & \cellcolor{gray!50}42.33 & \cellcolor{gray!50}39.18 & \cellcolor{gray!50}16.95 & \cellcolor{gray!50}46.00 & \cellcolor{gray!50}44.59 & \cellcolor{gray!50}43.54 & \cellcolor{gray!50}39.32 & \cellcolor{gray!50}16.77 & \cellcolor{gray!50}31.17 & \cellcolor{gray!50}30.84 & \cellcolor{gray!50}30.71 & \cellcolor{gray!50}27.94 & \cellcolor{gray!50}13.94 & \cellcolor{gray!50}32.17 & \cellcolor{gray!50}31.38 & \cellcolor{gray!50}31.16 & \cellcolor{gray!50}28.14 & \cellcolor{gray!50}14.31 \\ \bottomrule
\end{tabular}%
\begin{tablenotes}
\item[*] \hypertarget{tablenote:explanation} %
The highest metrics are colored with \colorbox{gray!50}{gray}. 
All metrics of UIED and Xianyu are lower than 0.01\% and omitted.\shuqing{Cite literature (similar findings). Notes locations. To main content.}
\end{tablenotes}
\end{threeparttable}
}
\vspace{-2em}
\end{table*}

\subsubsection{RQ1-1: Performance in terms of interactability}

As shown in Table~\ref{tab:model-performance-is}, \tool shows consistent performance in interactability detection across different IoU thresholds. Consider the generally best model \tool with Gemini, at the 0.75 threshold, 
the mAP peaks at 35.03\%, indicating a strong ability to detect \ivos accurately. However, as the IoU threshold increases, indicating a stricter criterion for object detection, there is a noticeable decrease in Precision, dropping to 5.64\% at the IoU of 0.95. 
This trend suggests that while the \tool is quite effective at a broader detection scope, its precision in highly specific object identification contexts is limited.
\tool demonstrates a balanced performance at lower IoU thresholds, with a peak F1 Score of 46.00\% at an IoU of 0.75. This balance 
is crucial for practical applications, as it indicates a well-rounded capability in both correctly identifying \ivos and minimizing false positives.

For the comparative evaluation, 
\tool generally demonstrates superior performance across IoU thresholds when compared with baseline models, especially on the genre split of the dataset, highlighting \tool's capability to handle various genres of apps. At an IoU threshold of 0.75, \tool achieves a Precision of 49.45\%, which is approximately 48.0\% higher than the best baseline, Faster-RCNN~\cite{paper:journals/pami/fasterrcnn}, at 33.42\%. This trend of superiority is more significant at higher IoU thresholds. At the stringent IoU of 0.95, \tool\ maintains a Precision of 19.02\%, outperforming the closest baseline, CenterNet2~\cite{paper:journals/corr/centernet2}, by over 321.7\%, a significant margin considering the complexity of XR environments.
In terms of Recall and F1 Score, \tool significantly outperforms other models by considerable margins. At an IoU of 0.75, \tool's Recall and F1 Score are 43\% and 46.00\%, respectively, surpassing CenterNet2~\cite{paper:journals/corr/centernet2}'s 30\% and 29.32\% by a notable margin of 43.3\% and 56.9\%, respectively. This indicates that \tool is not only accurate but also reliable in identifying \ivos, a key requirement in industrial XR apps.
Note that the direct prompt method on LMMs performs poorly. This is because although the LMM can identify the objects well, it can not effectively localize them, as discussed in~\cref{sec:orientor-motivating-example}. This result again reveals the inability of LMM alone to address the challenge in \ivo detection, highlighting \tool's enhancement for LMM on this task.

\subsubsection{RQ1-2: Performance in terms of semantics}

\tool's performance in semantics shows a similar but lower pattern compared to interactability. 
For \tool with Gemini, the mAP starts at a lower rate of 22.12\% at an IoU of 0.75 and follows a smooth decreasing trend with increasing IoU thresholds until 0.95, with a drop to 7.41\%. This pattern reflects the challenges inherent in semantic interpretation, especially under stringent detection criteria.

The Precision and Recall metrics provide further insights. While the Recall maintains at 50.00\%, the Precision starts at 23.72\% at an IoU of 0.75 and gradually decreases. This suggests that \tool maintains its ability to identify relevant semantic elements. However, \tool faces challenges in high precision scenarios, particularly as the criteria become more stringent.

As for the comparative evaluation in terms of semantics, the results in Table~\ref{tab:model-performance-semantics} reveal \tool's remarkable capability in semantic understanding. At an IoU of 0.75, \tool achieves a Precision of 23.72\% with Gemini on the genre split, which is significantly higher than the best baseline CenterNet2~\cite{paper:journals/corr/centernet2}'s 0.62\%, marking an increase of 3725.8\%. Similarly, at an IoU of 0.95, \tool's F1 Score of 14.31\% far surpasses the closest baseline, YOLO v8~\cite{software:yolov8}, which scores 0.40\%. This represents an increase of up to 3477.5\%, highlighting \tool's ability to understand complex semantic structures within XR environments. The direct prompting method continues to perform poorly in terms of semantics due to the aforementioned challenges that LMM alone can not effectively address. Note that the method directly prompting Gemini achieves a recall of 50\% on the genre split at 0.75 IoU threshold. This result suggests that LMM can perform well in some special settings.

Our analysis demonstrates that \tool significantly outperforms existing baseline models in 
both interactability and semantic understanding. The framework exhibits remarkable precision and recall rates, particularly in high IoU thresholds, which are critical for the nuanced and complex nature of industrial XR apps. 
This efficacy, especially in semantic understanding, underscores \tool's potential in revolutionizing \ivo detection in XR environments.

\subsubsection{RQ1-3: Performance in terms of context-sensitive interactability}

\begin{table}[t!]
\vspace{-1em}
\centering
\caption{Performance w.r.t. context-sensitive interactability, DL-based baselines are trained on our dataset\protect\hyperlink{tablenote:explanation}{\textsuperscript{*}}}
\label{tab:model-performance-context}
\vspace{-1em}
\resizebox{0.7\linewidth}{!}{%
\begin{tabular}{@{}c|ccc|ccc|ccc|ccc|ccc@{}}
\toprule
 & \multicolumn{3}{c|}{Faster-RCNN} & \multicolumn{3}{c|}{CenterNet2} & \multicolumn{3}{c|}{Yolo v8} & \multicolumn{3}{c|}{OmniParser} & \multicolumn{3}{c}{gpt-4o-2024-08-06} \\ 
IoU & P\% & R\% & F1\% & P\% & R\% & F1\% & P\% & R\% & F1\% & P\% & R\% & F1\% & P\% & R\% & F1\% \\ \midrule
0.75 & 2.44 & 0.14 & 0.27 & 4.27 & 0.27 & 0.51 & <0.01 & <0.01 & <0.01 & <0.01 & <0.01 & <0.01 & 2.44 & 0.07 & 0.13 \\
0.80 & 2.44 & 0.09 & 0.18 & 4.88 & 0.27 & 0.52 & <0.01 & <0.01 & <0.01 & <0.01 & <0.01 & <0.01 & 2.44 & 0.07 & 0.13 \\
0.85 & <0.01 & <0.01 & <0.01 & 4.88 & 0.23 & 0.43 & <0.01 & <0.01 & <0.01 & <0.01 & <0.01 & <0.01 & <0.01 & <0.01 & <0.01 \\
0.90 & <0.01 & <0.01 & <0.01 & 4.88 & 0.18 & 0.34 & <0.01 & <0.01 & <0.01 & <0.01 & <0.01 & <0.01 & <0.01 & <0.01 & <0.01 \\
0.95 & <0.01 & <0.01 & <0.01 & <0.01 & <0.01 & <0.01 & <0.01 & <0.01 & <0.01 & <0.01 & <0.01 & <0.01 & <0.01 & <0.01 & <0.01 \\ \midrule
 & \multicolumn{3}{c|}{Gemini 1.5 Pro} & \multicolumn{3}{c|}{\makecell[c]{{\textbf{\tool}} \\ \textbf{(gpt-4-vision-preview)}}} & \multicolumn{3}{c|}{\makecell[c]{{\textbf{\tool}} \\ \textbf{(gpt-4o-2024-08-06)}}} & \multicolumn{3}{c|}{\makecell[c]{{\textbf{\tool}} \\ \textbf{(Claude 3.5 Sonnet)}}} & \multicolumn{3}{c}{\makecell[c]{{\textbf{\tool}} \\ \textbf{(Gemini 1.5 Pro)}}} \\
IoU & P\% & R\% & F1\% & P\% & R\% & F1\% & P\% & R\% & F1\% & P\% & R\% & F1\% & P\% & R\% & F1\% \\ \midrule
0.75 & <0.01 & <0.01 & <0.01 & 44.72 & 6.66 & 10.77 & 67.70 & \cellcolor{gray!50}24.00 & \cellcolor{gray!50}32.51 & 62.74 & 18.54 & 26.07 & \cellcolor{gray!50}67.91 & 22.96 & 31.40 \\
0.80 & <0.01 & <0.01 & <0.01 & 44.57 & 6.32 & 10.41 & \cellcolor{gray!50}67.99 & \cellcolor{gray!50}23.36 & \cellcolor{gray!50}31.82 & 62.93 & 17.94 & 25.36 & 67.91 & 22.20 & 30.63 \\
0.85 & <0.01 & <0.01 & <0.01 & 45.02 & 6.27 & 10.40 & 67.82 & \cellcolor{gray!50}22.21 & \cellcolor{gray!50}30.62 & 64.17 & 16.84 & 24.27 & \cellcolor{gray!50}68.42 & 21.40 & 29.75 \\
0.90 & <0.01 & <0.01 & <0.01 & 45.24 & 5.37 & 9.26 & 68.93 & \cellcolor{gray!50}19.35 & \cellcolor{gray!50}27.75 & 63.96 & 14.84 & 21.84 & \cellcolor{gray!50}70.01 & 18.23 & 26.78 \\
0.95 & <0.01 & <0.01 & <0.01 & 26.83 & 1.62 & 3.03 & 47.56 & \cellcolor{gray!50}8.99 & \cellcolor{gray!50}13.59 & 47.07 & 7.47 & 11.37 & \cellcolor{gray!50}53.35 & 7.01 & 11.76 \\ \bottomrule
\end{tabular}
}
\vspace{-1em}
\end{table}

As shown in table \ref{tab:model-performance-context}, \tool shows a promising performance to distinguish \ivos' interactability in different contexts. \tool with Gemini performs better in terms of Precision, which achieves 67.91\% at the 0.75 IoU threshold and maintains superior performance over 65\% as the threshold increases, until at IoU threshold of 0.95, dropping to 53.35\% which is still relatively high. \tool with GPT-4o shows a better performance in terms of Recall and F1 score, achieving 24.00\% and 32.51\% at the 0.75 IoU threshold, demonstrating its better balance between Precision and Recall. Regarding the comparison with baselines, \tool shows superior performance against baseline methods across different IoU thresholds. Consider the generally better model, \tool with GPT-4o, at the lower IoU threshold of 0.75, \tool achieves an F1 score of 32.51\%, surpassing the best baseline CenterNet2~\cite{paper:journals/corr/centernet2} by 6274.5\%. At the most stringent constraint of 0.95 IoU threshold, \tool's F1 score maintains over 10\% while all metrics of baseline models are approximately 0, demonstrating \tool's significant advantage in understanding context-sensitive \ivos' interactability against baseline models.

\subsection{RQ2: Ablation Study}

\begin{table*}[t!]
\centering
\caption{Contributions of different components\protect\hyperlink{tablenote:explanation}{\textsuperscript{*}}}
\vspace{-1em}
\label{tab:model-performance-ablation}
\resizebox{0.85\linewidth}{!}{
\begin{threeparttable}
\setlength{\tabcolsep}{2pt}
\begin{tabular}{@{}l|rrrrrr|rrrrrr|rrrrrr|rrrrrr@{}}
\toprule
{} & \multicolumn{6}{c|}{\tool} & \multicolumn{6}{c|}{\textit{w/o Context Comprehension}} & \multicolumn{6}{c|}{\textit{w/o Reflection-Directed Loop}} & \multicolumn{6}{c}{\textit{w/o Interactability Classification}}\\ 
{} & \multicolumn{3}{c|}{Interactability} & \multicolumn{3}{c|}{Semantic} & \multicolumn{3}{c|}{Interactability} & \multicolumn{3}{c|}{Semantic} & \multicolumn{3}{c|}{Interactability} & \multicolumn{3}{c|}{Semantic} & \multicolumn{3}{c|}{Interactability} & \multicolumn{3}{c}{Semantic} \\
IoU & P\% & R\% & F1\% & P\% & R\% & F1\% & P\% & R\% & F1\% & P\% & R\% & F1\% & P\% & R\% & F1\% & P\% & R\% & F1\% & P\% & R\% & F1\% & P\% & R\% & F1\% \\
\midrule

0.75 & \cellcolor{gray!50}47.14 & \cellcolor{gray!50}43.00 & \cellcolor{gray!50}44.97 & \cellcolor{gray!50}21.36 & \cellcolor{gray!50}50.00 & \cellcolor{gray!50}29.93 & 38.27 & 39.00 & 38.63 & 18.14 & \cellcolor{gray!50}50.00 & 26.62 & 38.51 & 36.00 & 37.21 & 18.85 & \cellcolor{gray!50}50.00 & 27.38 & 31.40 & 29.00 & 30.15 & 16.78 & \cellcolor{gray!50}50.00 & 25.13 \\
0.80 & \cellcolor{gray!50}45.95 & \cellcolor{gray!50}42.00 & \cellcolor{gray!50}43.88 & \cellcolor{gray!50}20.82 & \cellcolor{gray!50}50.00 & \cellcolor{gray!50}29.40 & 35.80 & 39.00 & 37.33 & 17.71 & \cellcolor{gray!50}50.00 & 26.15 & 36.18 & 36.00 & 36.09 & 18.34 & \cellcolor{gray!50}50.00 & 26.83 & 29.63 & 29.00 & 29.31 & 16.47 & \cellcolor{gray!50}50.00 & 24.78 \\
0.85 & \cellcolor{gray!50}45.37 & \cellcolor{gray!50}41.00 & \cellcolor{gray!50}43.08 & \cellcolor{gray!50}20.64 & \cellcolor{gray!50}50.00 & \cellcolor{gray!50}29.22 & 34.90 & 38.00 & 36.38 & 17.13 & \cellcolor{gray!50}50.00 & 25.51 & 36.36 & 34.00 & 35.14 & 17.57 & 50.00 & 26.01 & 29.91 & 27.00 & 28.38 & 18.90 & 33.00 & 24.04 \\
0.90 & \cellcolor{gray!50}41.76 & \cellcolor{gray!50}37.00 & \cellcolor{gray!50}39.23 & \cellcolor{gray!50}17.69 & \cellcolor{gray!50}50.00 & \cellcolor{gray!50}26.14 & 33.00 & 33.00 & 33.00 & 17.13 & 33.00 & 22.55 & 32.87 & 31.00 & 31.91 & 14.51 & \cellcolor{gray!50}50.00 & 22.49 & 26.77 & 25.00 & 25.86 & 16.46 & 33.00 & 21.97 \\
0.95 & \cellcolor{gray!50}19.25 & \cellcolor{gray!50}16.00 & \cellcolor{gray!50}17.47 & \cellcolor{gray!50}8.18 & \cellcolor{gray!50}50.00 & \cellcolor{gray!50}14.06 & 13.80 & 14.00 & 13.90 & 6.86 & 33.00 & 11.36 & 13.91 & 13.00 & 13.44 & 6.82 & 33.00 & 11.30 & 11.58 & 10.00 & 10.73 & 7.22 & 16.00 & 9.95 \\

\bottomrule
\end{tabular}
\end{threeparttable}
}
\end{table*}

Table~\ref{tab:model-performance-ablation} shows the performance of \tool and its three variations. 
We remove the corresponding components of each variation to demonstrate their neceesity.

As can be seen in Table~\ref{tab:model-performance-ablation}, \tool outperforms all its variants significantly, demonstrating the necessity of the designed pipeline. %
For the \textit{Context Comprehension} component, its absence results in noticeable performance degradation. Compared to \tool, the F1 score drops by 14.10\% (38.63\% vs. 44.97\%) and 11.07\% (26.62\% vs. 29.93\%) at 0.75 IoU threshold for the interactability and semantic tasks, respectively. The gap grows even wider as the IoU threshold increases, as such a trend also appears in other metrics, underscoring the importance of semantics context comprehension in improving \tool's ability to correctly detect \ivos across varying IoU thresholds.
Replacing \textit{\ivo Feature Mining \& Referring} components results in a significant weakening of the performance, with an average decline of 19.04\% and 12.37\% on F1 Score when distinguishing \ivos' interactability and semantics, suggesting the critical role of the Candidate \ivo Referring components.
As for the \textit{Interactability Classification} component, removing this component leads to the most significant performance degradation. Compared to \tool, the F1 score decreases by 34.60\% and 18.94\% on average in interactability and semantics, respectively. The experimental results show the necessity of each component in \tool, highlighting their critical role in accurately detecting \ivos.

\subsection{RQ3: Effectiveness for boosting automated testing\shuqing{More discussion}}
\label{sec:rq-usefulness}

\begin{table}[t!]
\centering
\caption{Average testing result by apps with different categories. Categories with \# of app < 10 are omitted}
\label{tab:test-by-category}
\resizebox{0.7\columnwidth}{!}{%
\begin{tabular}{@{}lrrrrr@{}}
\toprule
\textbf{App Category} & \textbf{\# App} & \textbf{\# GUI Scene} & \textbf{Random \ivo Cov.} & \textbf{Guided \ivo Cov.} & \textbf{Improved} \\
\midrule
All & 77 & 936 & 0.65 & 0.83 & +26.91\% \\ \midrule
\rowcolor{mygray} Singleplayer & 43 & 640 & 0.45 & 0.60 & +32.42\% \\
Casual & 46 & 600 & 0.42 & 0.52 & +24.75\% \\
\rowcolor{mygray}Simulation & 30 & 521 & 0.37 & 0.49 & +31.05\% \\
3D & 24 & 471 & 0.35 & 0.47 & +33.43\% \\
\rowcolor{mygray}Immersive Sim & 13 & 374 & 0.28 & 0.38 & +35.05\% \\
First-Person & 28 & 402 & 0.29 & 0.36 & +25.61\% \\
\rowcolor{mygray}Puzzle & 14 & 339 & 0.25 & 0.34 & +36.60\% \\
Free to Play & 36 & 393 & 0.27 & 0.33 & +20.64\% \\
\rowcolor{mygray}Early Access & 11 & 319 & 0.24 & 0.32 & +32.52\% \\
Indie & 38 & 407 & 0.27 & 0.32 & +19.88\% \\
\rowcolor{mygray}Colorful & 13 & 318 & 0.24 & 0.32 & +35.39\% \\
6DOF & 14 & 295 & 0.21 & 0.27 & +29.22\% \\
\rowcolor{mygray}Education & 12 & 292 & 0.21 & 0.27 & +31.40\% \\
Atmospheric & 24 & 267 & 0.18 & 0.21 & +15.05\% \\
\rowcolor{mygray}Adventure & 23 & 224 & 0.15 & 0.19 & +26.29\% \\
Action & 24 & 220 & 0.15 & 0.17 & +13.06\% \\
\rowcolor{mygray}Physics & 14 & 178 & 0.12 & 0.15 & +21.87\% \\
Exploration & 14 & 181 & 0.12 & 0.15 & +21.17\% \\
\rowcolor{mygray}Arcade & 17 & 163 & 0.11 & 0.13 & +25.59\% \\
Funny & 10 & 93 & 0.05 & 0.07 & +34.46\% \\
\rowcolor{mygray}Strategy & 11 & 87 & 0.06 & 0.07 & +13.52\% \\
Sci-fi & 11 & 65 & 0.04 & 0.04 & +8.84\% \\
\rowcolor{mygray}Multiplayer & 11 & 56 & 0.03 & 0.03 & -11.37\% \\ \midrule
Mean & 20 & 300 & 0.21 & 0.27 & +24.19\% \\
\bottomrule
\end{tabular}
}
\end{table}

\begin{wrapfigure}{r}{0.53\linewidth}
    \centering 
    \vspace{-2em}
	\subfigure[Effective interaction count]{
		\label{fig:effective-interaction-count}
		\includegraphics[width=0.48\linewidth]{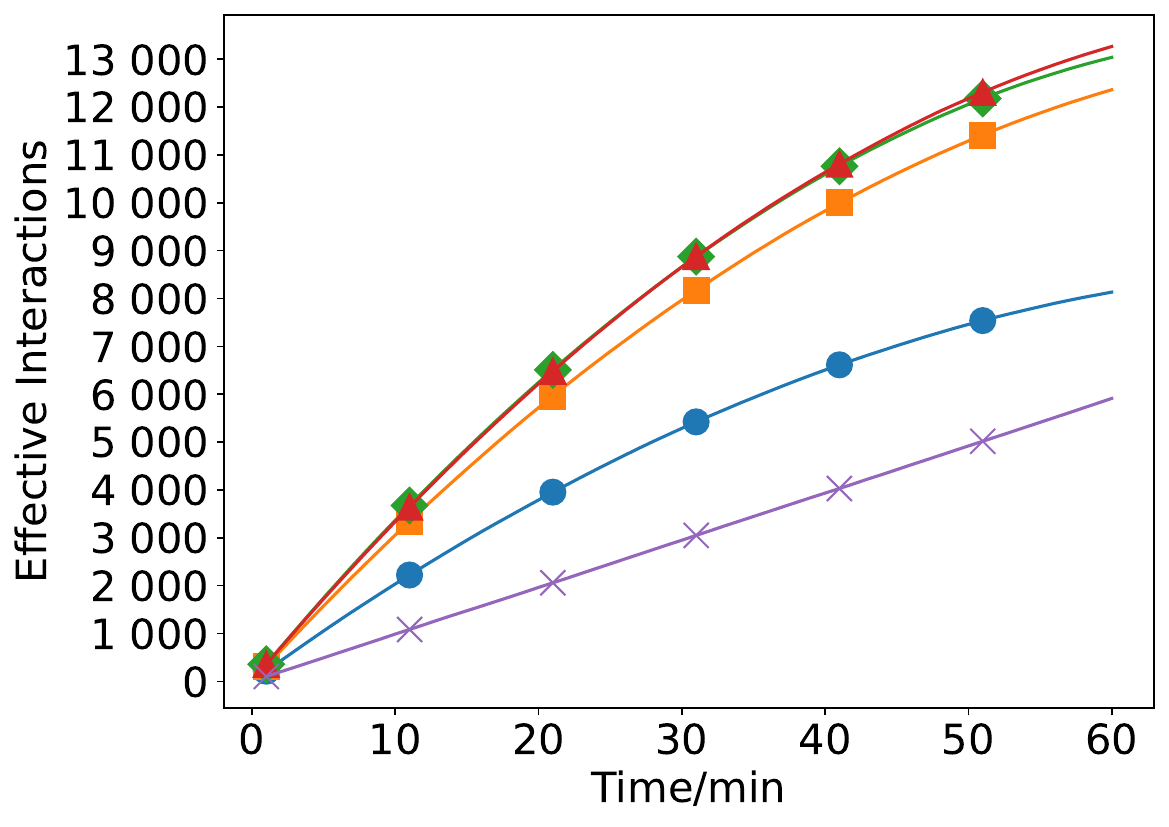}}
	\subfigure[Coverage rate]{
		\label{fig:coverage-rate}
		\includegraphics[width=0.45\linewidth]{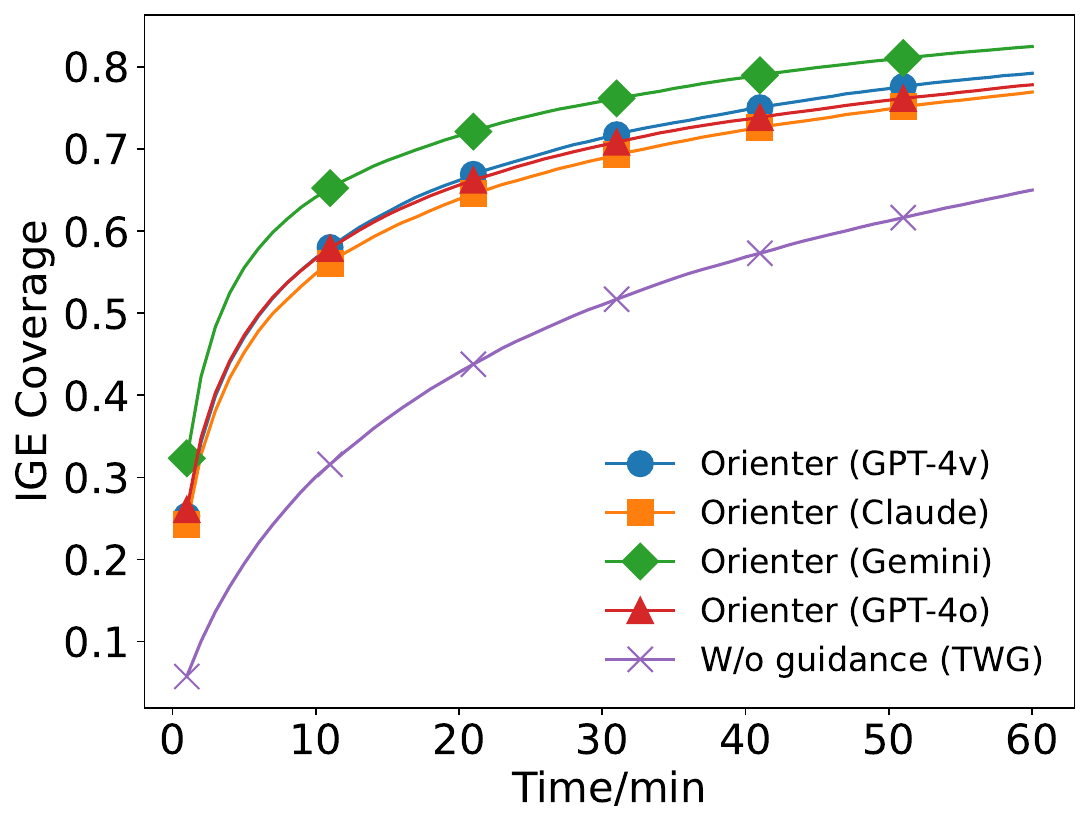}}
    \vspace{-1.5em}
    \caption{Performance of testing over time across strategies}
    \label{fig:interaction-gen-comparison}
\end{wrapfigure}

As shown in Figure~\ref{fig:interaction-gen-comparison}, the testing guided by \tool outperforms the random testing with a significant gap. We select the best model, \tool with Gemini for comparison.
The number of effective interactions in guided testing increases notably faster than in random generation. At the end of the testing, the guided one's average number of effective interactions reaches 13,038.03 on 936 GUI Scenes in total, surpassing the random testing's 5,911.53 by a remarkable 120.6\%. 
At 10 minutes, the \ivo coverage of the guided testing already reaches 0.64, surpassing the random testing's 0.30 by 113.3\%. The guided testing eventually reaches a coverage of 0.83, 27.7\% higher than the random testing's 0.65. 

\begin{figure}[t!]
 	\centering 
 	\includegraphics[width=0.6\linewidth]{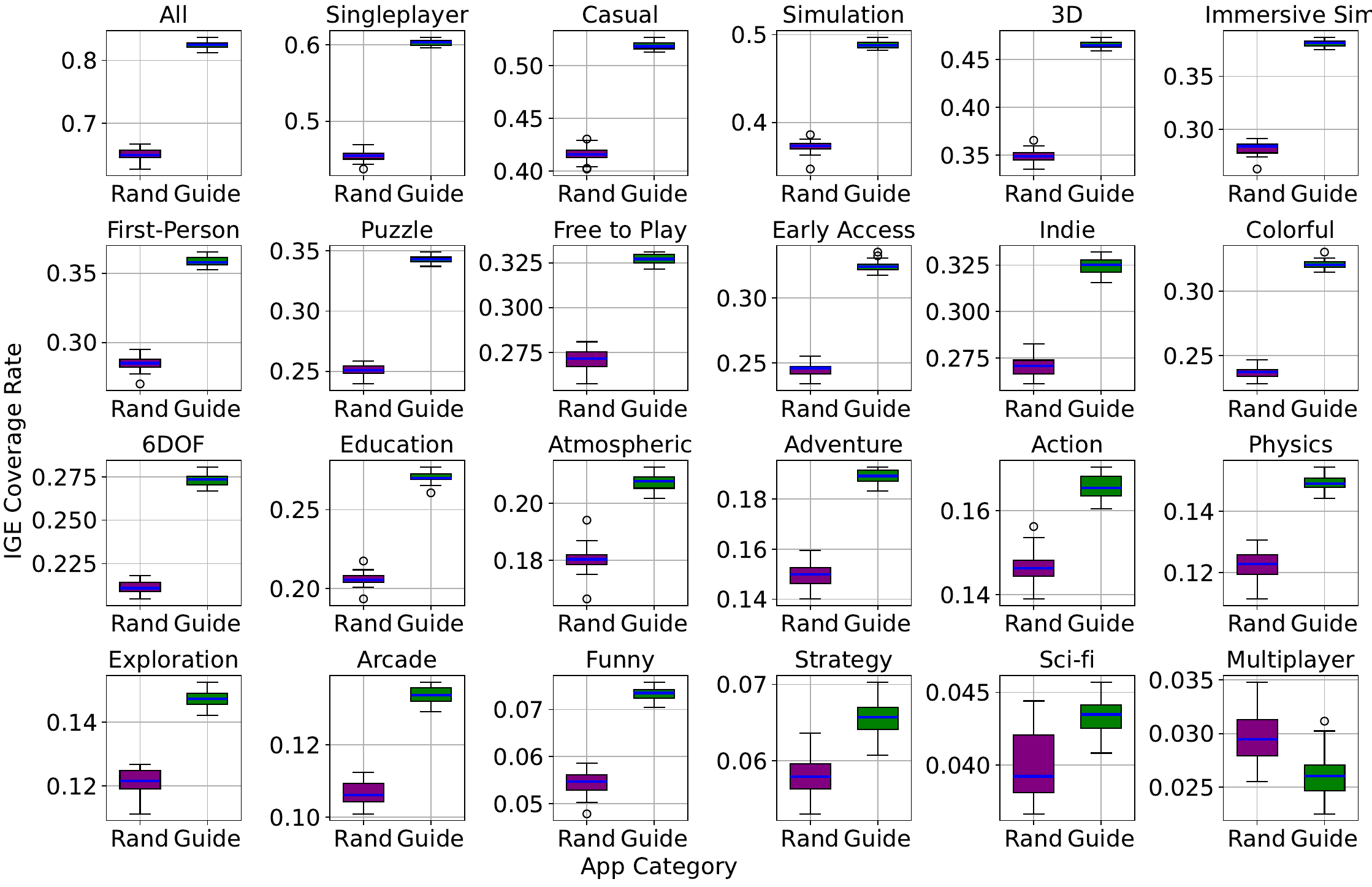} 
        \vspace{-1em}
 	\caption{Coverage rate on apps in different categories. Categories with \# of app < 10 are omitted}
 	\label{fig:test-coverage-rate-by-category}
\end{figure}

We further investigate the performance of guided and non-guided testing agents on apps with different categories. We select 23 community-generated categories with no less than 10 apps (except for the category "VR") for category-wise evaluation in terms of \ivo coverage, covering 76 of the 77 AUTs. 
Table~\ref{tab:test-by-category} and Figure~\ref{fig:test-coverage-rate-by-category}
intuitively reveal the benefits of the guidance from \tool. 
The testing agent with the guidance of \tool surpasses the one without guidance in most app categorys, achieving improvements by up to 36.60\% (0.34 vs. 0.25) in the category "Puzzle". The guided agent achieves the highest \ivo coverage of 0.60 in the category "Singleplayer", surpassing the random agent by 32.42\%. 
In contraction, the guided testing agent shows a worse yet still very close \ivo coverage compared with the random testing agent in the "Multiplayer" category, both achieving a coverage at around 0.03. Such reduction suggests that the bias introduced by the external guidance could also lead to performance degradation for testing agents, highlighting the necessity of "thinking outside the box" to explore the GUI scene aside from the guidance to some extent. The large gap between the two opposite categories also reflects the huge difference between different categories of apps.
In addition, the guided testing agent generally performs more consistently than the random testing agent between testing rounds.

These results demonstrate the capability of \tool to boost the effectiveness and efficiency of XR app testing. Apart from automated XR app testing, other downstream software engineering problems like UI to code, video-based or screenshot-based testing, etc., can also benefit from the ability of \tool to effective and efficient \ivo detection.
\shuqing{Summarize discussions on usefulness to here.}

\todoaftersub{RQ: Different category}

\todoaftersub{Cannot directly use llm and LMM. No multimodel and exact output.}

\todoaftersub{Huge search space.}

\label{sec:usefulness}

\section{Threats to Validity}

\textbf{Internal Validity.}
An internal threat to validity is related to the annotators' bias and subjective interpretation.
Given that the dataset was annotated by 13 annotators, there's potential bias based on their personal experience and interpretation of interactability in XR apps. This could impact the consistency and accuracy of the annotations.
The process of labeling semantics of \ivos is subjective and dependent on the annotator's understanding and perception, which may not always align with the intended use or perception in different user demographics.
To mitigate it, we construct a detailed and objective data collection and annotation process. We spend at least two hours for each annotator to make sure they understand all steps.
Another threat is model bias in pretrained models,
given that \tool relies on pretrained LMMs, there is a risk that biases inherent in these models could affect the accuracy of \ivo detection. If these models were trained on data that was not representative of diverse contexts, their predictions could be skewed.
To mitigate this threat, we choose the most powerful and representative models, which are trained on large corpus and which are claimed that the team has reduced model bias to some level.

\textbf{External Validity.}
One threat to external validity is the generalizability across XR apps. The diversity in XR apps poses a challenge to generalizability. \tool's effectiveness demonstrated in specific scenarios may not uniformly apply across different XR environments, particularly those with unique interaction paradigms or novel GUI element types.
To mitigate this threat, we sample popular and complicated XR apps, crossing a large portion of app categories.

\todoaftersub{Data, yolo, foreground/background}
\section{Related Work}

\subsection{IGE Detection\shuqing{UI analysis to IGE detection.}}
\label{subsec:related-work-ui-analysis}

Empirical studies have been conducted to give insights into IGE detection~\cite{paper:conf/fse/YeCX0HCXZ21}, e.g.,  Chen et al.~\cite{paper:conf/fse/ChenXXCXZ020/od} conducted a systematic large-scale empirical study on GUI element detection methods. 
Previous works have adapted both old-fashioned methods ~\cite{paper:conf/uist/YehCM09,paper:conf/icse/BaoLXWZ15,paper:conf/chi/DixonF10} and deep learning models~\cite{paper:conf/sigsoft/WangLGLXD23, paper:conf/kbse/QianMLC22, paper:conf/icse/WuYCXHHMZ23} to IGE detection. 
White et al.~\cite{paper:conf/issta/WhiteFB19} replaced methods that involve GUI APIs with machine learning techniques to find interactable elements in GUI images. 
Xie et al.~\cite{paper:conf/fse/XieFXCC20} proposed UIED, which combines old-fashioned computer vision approaches and deep learning models to detect components on complex GUI images. 
For game apps, 
Ye et al. ~\cite{paper:conf/fse/YeCX0HCXZ21} conducted an empirical study of clickable GUI element detection on mobile games and constructed the first clickable GUI element detection benchmark.  
Wu et al.~\cite{paper:conf/icse/WuYCXHHMZ23} constructed a larger and more precise GUI dataset for mobile games and evaluated seven IGE detection techniques on it.
The industry also applies IGE detection techniques to solve business problems like GUI testing~\cite{paper:conf/icse/RanLLWMWJCTX22, paper:conf/sigsoft/WangLGLXD23, paper:conf/kbse/QianMLC22}. 
Ran et al.~\cite{paper:conf/icse/RanLLWMWJCTX22} reported their experiences and lessons learned developing and deploying VTest, an automated visual testing framework for smartphones that leverages only GUI images. 
Wang et al. ~\cite{paper:conf/sigsoft/WangLGLXD23} proposed iExplorerGame, a unified framework for black-box mini-game testing that combines deep learning object detection and edge aggregation-based segmentation to guide the testing. 
Qian et al. ~\cite{paper:conf/kbse/QianMLC22} proposed a fast OCR-based widget localization technique Label Text Screening\shuqing{Revise.}, which accelerates the OCR widget detection by analyzing and leveraging the feature of texts in widgets.
Research on UI analysis is going in-depth and reaching fields with complex GUI-like game apps. However, existing methods are limited to common \ivo with finite predefined categories, which fail to tackle the open vocabulary challenge.

\subsection{VR/AR Testing}

VR/AR (XR) apps provide an immersive experience to users, involving various procedures like device tracking and rendering. To ensure users' experience, there are several works conducting empirical studies on XR software testing~\cite{paper:conf/svr/AndradeND19}. 
Andrade et al.~\cite{paper:conf/svr/AndradeND19} compared open-source VR projects with non-VR projects to point out the necessity of performing testing on VR apps. 
Rzig et al.~\cite{paper:vr-testing-empirical} performed a large-scale empirical study on software testing practices of open-source VR projects.
Several works have proposed approaches to facilitate XR testing. Qin and Hassan~\cite{paper:conf/kbse/QinH22} proposed DyTRec, which provides developers with recommendations on which codes should be tested by actually running the app. 
Souza et al.~\cite{paper:journals/stvr/SouzaND18} proposed VR-ReST, a tool designed to assist with requirements specification and test data generation for VR apps. 
Rafi et al.~\cite{paper:predart} proposed PredART which can be used as the test oracle when checking the placement of virtual objects in AR. 
Wang~\cite{paper:conf/icse/Wang22} proposed VRTest that extracts information from the VR scene and automatically explores the scene and interacts with the objects by controlling the camera. Andrade et al.~\cite{paper:journals/stvr/AndradeFM23} proposed a novel approach that combines metamorphic testing, agent-based testing, and machine learning to test VR apps.
Recently, Li et al.~\cite{paper:stereoid} proposed StereoID to automatically detect stereoscopic visual inconsistencies in VR apps.
\shuqing{StereoID paper. Done.}

\subsection{Exploratory Studies on VR/AR Apps}

Previous works have conducted several empirical studies to understand VR/AR (XR) apps from different perspectives. Rodriguez and Wang~\cite{paper:os-vr-rodriguez17} analyzed the growing trends, amount of developers, popular topics, and common files in open-source VR software projects. Li et al.~\cite{paper:issre-webxr-bugs} conducted an empirical study of bugs in web-based XR projects to understand their symptoms and uniqueness. Adam et al.~\cite{paper:sec-interview-adams18} conducted interviews with VR users and developers and surveyed their concerns about the security and privacy of VR apps. Nusrat et al.~\cite{paper:vr-performance-optimization} systematically performed an empirical study of performance optimization in various VR projects. 
Li et al.~\cite{paper:vr-software-quality} modelled the software quality attributes and key influencing factors of VR applications from the users' perspectives.
Guo et al.~\cite{paper:conf/icse/GuoDLZXH24} developed a security and privacy assessment tool and conducted an empirical study on Oculus VR apps.

\section{Conclusion}
In this paper, we analyze the key challenges in interactable GUI element (\ivo) detection for Extended Reality (XR) apps.
We propose \tool, the first zero-shot, context-sensitive IGE detection framework for XR apps. 
Extensive experiments have been conducted to verify the effectiveness of the proposed IGE detection framework. The results demonstrate that Orienter is more effective than the state-of-the-art GUI element detection approaches,
Experiments also illustrate that \tool is beneficial for boosting automated GUI testing.
Alongside developing \tool, we have also created the first dedicated dataset for \ivo detection in XR environments, both of which are publicly available to spur further research in this field. 

\balance
\bibliographystyle{ACM-Reference-Format}
\bibliography{acmart}

\end{document}